\documentclass[iop]{emulateapj}
\usepackage{appendix,natbib}
\usepackage[pdfpagemode=UseNone,pdfstartview=FitH,backref,colorlinks=true,citecolor=blue,linkcolor=blue,urlcolor=blue]{hyperref}
\usepackage[all]{hypcap}

\newcommand{\halpha}{H\ensuremath{\alpha}}
\newcommand{\hbeta}{H\ensuremath{\beta}}
\newcommand{\ebmv}{$E(B-V)$}
\def\msun{{\rm\,M_\odot}}

\begin{document}

\title{ The MOSDEF Survey: Dissecting the star-formation rate vs. stellar mass relation using {\halpha} and {\hbeta} emission lines at $\lowercase{z}\sim 2$}

\author{\sc Irene Shivaei\altaffilmark{1,5}, Naveen A. Reddy\altaffilmark{1,6}, Alice E. Shapley\altaffilmark{2}, Mariska Kriek\altaffilmark{3}, Brian Siana\altaffilmark{1}, Bahram Mobasher\altaffilmark{1}, Alison L. Coil\altaffilmark{4}, William R. Freeman\altaffilmark{1}, Ryan Sanders\altaffilmark{2}, Sedona H. Price\altaffilmark{3}, Laura de Groot\altaffilmark{1}, Mojegan Azadi\altaffilmark{4}}

\altaffiltext{1}{Department of Physics \& Astronomy, University of California, Riverside, CA 92521, USA}
\altaffiltext{2}{Department of Physics \& Astronomy, University of California, Los Angeles, CA 90095, USA}
\altaffiltext{3}{Astronomy Department, University of California, Berkeley, CA 94720, USA}
\altaffiltext{4}{Center for Astrophysics and Space Sciences, University of California, San Diego, La Jolla, CA 92093, USA}
\altaffiltext{5}{NSF Graduate Research Fellow}
\altaffiltext{6}{Alfred P. Sloan Research Fellow}

\slugcomment{Accepted for publication in ApJ}

\begin{abstract}
We present results on the star-formation rate (SFR) versus stellar mass ($M_*$) relation (i.e., the ``main sequence'') among star-forming galaxies at 1.37\,$\leq$\,$z$\,$\leq$\,2.61 using the MOSFIRE Deep Evolution Field (MOSDEF) survey. Based on a sample of 261 galaxies with {\halpha} and {\hbeta} spectroscopy, we have estimated robust dust-corrected instantaneous SFRs over a large range in $M_*$ ($\sim\,10^{9.5}-10^{11.5}\,\msun$). We find a correlation between log(SFR({\halpha})) and log($M_*$) with a slope of 0.65\,$\pm$\,0.08 (0.58\,$\pm$\,0.10) at 1.4\,$<$\,$z$\,$<$\,2.6 (2.1\,$<$\,$z$\,$<$\,2.6). We find that different assumptions for the dust correction, such as using the color-excess of the stellar continuum to correct the nebular lines, sample selection biases against red star-forming galaxies, and not accounting for Balmer absorption can yield steeper slopes of the log(SFR)-log($M_*$) relation. Our sample is immune from these biases as it is rest-frame optically selected, {\halpha} and {\hbeta} are corrected for Balmer absorption, and the {\halpha} luminosity is dust-corrected using the nebular color-excess computed from the Balmer decrement. The scatter of the log(SFR({\halpha}))-log($M_*$) relation, after accounting for the measurement uncertainties, is 0.31\,dex at 2.1\,$<$\,$z$\,$<$\,2.6, which is 0.05\,dex larger than the scatter in log(SFR(UV))-log($M_*$). Based on comparisons to a simulated SFR-$M_*$ relation with some intrinsic scatter, we argue that in the absence of direct measurements of galaxy-to-galaxy variations in the attenuation/extinction curves and the IMF, one cannot use the difference in the scatter of the SFR({\halpha})- and SFR(UV)-$M_*$ relations to constrain the stochasticity of star formation in high-redshift galaxies.

\end{abstract}
\maketitle

\section{Introduction}

Star-forming galaxies show a relatively tight correlation between their star-formation rate (SFR) and stellar mass. This relation, commonly known as the star-forming {\em ``main sequence''} \citep{noeske07a}, has been intensively studied over the past decade \citep[e.g., at $z\sim 2$,][]{reddy06a,daddi07a,pannella09,rodighiero11,whitaker12b,wuyts11b,reddy12b,sawicki12,schreiber14,whitaker14b,atek14,rodighiero14}. The main sequence is generally assumed to be a linear relation between log(SFR) and log($M_*$) with an associated scatter due to observational uncertainties and intrinsic scatter. 
The slope of the relation tells us about star formation efficiency as a function of stellar mass \citep{whitaker14b,genzel15}.
The intrinsic scatter of the log(SFR)-log($M_*$) relation is predicted to be due to variations in the gas accretion histories of different galaxies \citep{dutton10a} and indicates the level of burstiness in star formation history.
The common physical interpretation of the tightness of the SFR-$M_*$ relation is that the bulk of the stellar mass in star-forming galaxies is built in a relatively steady process, as opposed to a rapid starburst mode, as occurs in major mergers. The merger-driven starburst galaxies with high specific SFRs tend to lie above the main sequence relation \citep{rodighiero11,atek14} and the quiescent galaxies that have little ongoing star formation, identified by their red near-infrared colors \citep{williams09,nayyeri14}, populate a region below the sequence \citep{noeske07a,wuyts11b}.

It has been argued that the majority of galaxies spend most of their lifetime on the main sequence \citep{noeske07a}. Therefore, studying the parameters of the main sequence and comparing them with the galaxy formation models can shed light on our understanding of processes that govern galaxy evolution, such as stellar and AGN feedback \citep{kannan14,torrey14,sparre15}, gas accretion rates \citep{dutton10a}, and gas inflows and outflows \citep{dave11a}. The SFR-$M_*$ relation can be used as a test for galaxy evolution simulations to see how well the simulations can reproduce the observed universe \citep[e.g.,][]{behroozi13b}.

Determining the slope and intrinsic scatter of the log(SFR)-log($M_*$) relation can be hindered by sample selection effects, the diagnostic used to infer the SFR, and the method used to correct observed SFR for dust attenuation \citep[for a summary see,][]{speagle14}. The UV-selected samples and the UV-inferred SFRs are biased against massive and dusty galaxies where the bulk of star formation is obscured and the UV slope is decoupled from extinction. Moreover, as SED- and UV-based SFRs are dependent on the same stellar population models used to infer the stellar masses, the scatter in the log(SFR)-log($M_*$) relation may be underestimated when using SED or UV based SFR tracers \citep{reddy12b}. In order to overcome this issue, and more robustly measure SFRs for dustier galaxies, some studies have used SFR(IR)+SFR(UV) as a proxy for bolometric SFR \citep[e.g.,][]{reddy06a,schreiber14,whitaker14b}. However, estimating the SFR(IR) based solely on {\em Spitzer}/MIPS 24$\mu$m imaging has its shortcomings and could artificially tighten the main sequence if the scatter in L(IR) at a given L($8\mu$m) luminosity is not taken into account \citep[e.g.,][]{utomo14,hayward14}.

We improve upon previous high-redshift studies by utilizing a representative sample of galaxies with measurements of the {\halpha} and {\hbeta} emission lines. These lines offer the advantage of probing star formation on shorter timescales than the UV and IR, they are largely independent of the stellar population, and thus can be used to more accurately assess the intrinsic scatter in the log(SFR)-log($M_*$) relation. 

The star-formation activity was at its peak at $z\sim 2$ \citep{reddy09,shapley11}, making that redshift a critical epoch to study the evolution of galaxies. Obtaining rest-frame optical spectra of large samples of galaxies at $z\sim 2$ has been challenging until recently due to the high terrestrial background in the near-IR and the lack of multi-object near-IR spectrographs on 10-meter-class telescopes. The newly-commissioned MOSFIRE spectrograph \citep{mclean12} on the 10\,m Keck I telescope has enabled us to build a large sample of galaxies at $z\sim 2$ with coverage in both the {\halpha} and {\hbeta} lines.
In this study, we use a large sample of 261 rest-frame optically selected and spectroscopically confirmed galaxies at $1.4\,z\,2.6$, which was obtained as part of the MOSFIRE Deep Evolution Field (MOSDEF) survey \citep{kriek15}. To study the SFR-$M_*$ relation, we use SFRs based on the {\halpha} luminosity that are robustly corrected for dust attenuation using the Balmer decrement (L({\halpha})/L({\hbeta})). We demonstrate how different SFR diagnostics and dust correction recipes can alter the slope and scatter of the log(SFR)-log($M_*$) relation, and thus potentially affect our conclusions regarding the way in which galaxies build up their stellar mass at high redshift.

The outline of this paper is as follows. In Section~\ref{sec:sample}, we present our sample properties and our measurements including SFRs and stellar masses. In Section~\ref{sec:ms_analysis}, we describe the methodology we used to derive the log(SFR)-log($M_*$) relation parameters. We compare the scatter of the log(SFR)-log($M_*$) relation for different SFR indicators in Section~\ref{sec:scatter}. In Section~\ref{sec:slope}, we discuss the slope of the log(SFR)-log($M_*$) relation and the potential biases arising from sample selection, dust correction, and Balmer absorption correction. We compare our findings to previous studies in Section~\ref{sec:compare}, and finally, the results are summarized in Section~\ref{sec:summary}.
Throughout this paper, a \citet{chabrier03} Initial Mass Function (IMF) is assumed and a cosmology with H$_0$ = 70 km s$^{-1}$ Mpc$^{-1}$, $\Omega_{\Lambda}$ = 0.7, and $\Omega_m$ = 0.3 is adopted.
 
\section{Sample and Measurements}
\label{sec:sample}
\subsection{The MOSDEF Survey}

The MOSDEF survey is a large multi-year survey with the MOSFIRE multi-object spectrometer on the Keck I telescope \citep{mclean12}. The aim of the survey is to obtain rest-frame optical spectra of $\sim 1500$ {\em H}-selected galaxies to study their stellar, gaseous, metal, dust, and black hole content. The MOSDEF observations are conducted in five fields of the Cosmic Assembly Near-IR Deep Extragalactic Legacy Survey \citep[CANDELS;][]{koekemoer11,grogin11}, consisting of AEGIS, COSMOS, GOODS-N, GOODS-S, and UDS, targeting three redshift bins: $1.37\leq z\leq 1.70$, $2.09\leq z\leq 2.61$, and $2.95\leq z\leq 3.80$.
Targets for spectroscopy are prioritized by their {\em H}-band magnitude (using the 3D-HST catalogs, \citealt{skelton14}) and the availability of spectroscopic, grism, and photometric redshifts, down to {\em H} = 24.0, 24.5, and 25.0 magnitude, for each redshift bin respectively.

Details of the survey strategy, observations, data reduction, and characteristics of the full galaxy sample are described in \citet{kriek15}. In this study, we use the data accumulated from the first two years of the survey.

\subsection{{\halpha} Sample}
\label{sec:hasample}

Out of the full MOSDEF sample we selected a subsample of 342 galaxies that have secure redshifts and coverage of both the {\halpha} and {\hbeta} lines. These criteria limited our sample to the low and middle redshift bins that correspond to $1.37\leq z\leq 1.70$ and $2.09\leq z\leq 2.61$, respectively. There were 49 AGN that were identified and removed from the sample based on their IR and X-ray properties, as well as high [N{\sc ii}]/{\halpha} ratios ([N{\sc ii}]/{\halpha}$>0.5$, \citealt{coil15}). 
In order to have secure SFR measurements, we further limited our sample to objects where the {\halpha} line was detected with signal-to-noise (S/N) $> 3$. This selection resulted in a sample of 264 objects. 

Among the {\halpha}-detected galaxies, the {\hbeta} fluxes of 19\% (48 out of 264) were below $3\sigma$. Out of these 48 {\hbeta}-undetected objects, 40 (83\%) had a bright sky line close to the {\hbeta} line, while for the detected {\hbeta} sample the fraction was only 34\%. The higher frequency of skyline contamination in the {\hbeta}-undetected sample contributes to the lower {\hbeta} detection rate among these galaxies.

\begin{figure}[tbp]
	\begin{center}
		\includegraphics[width=.4\textwidth]{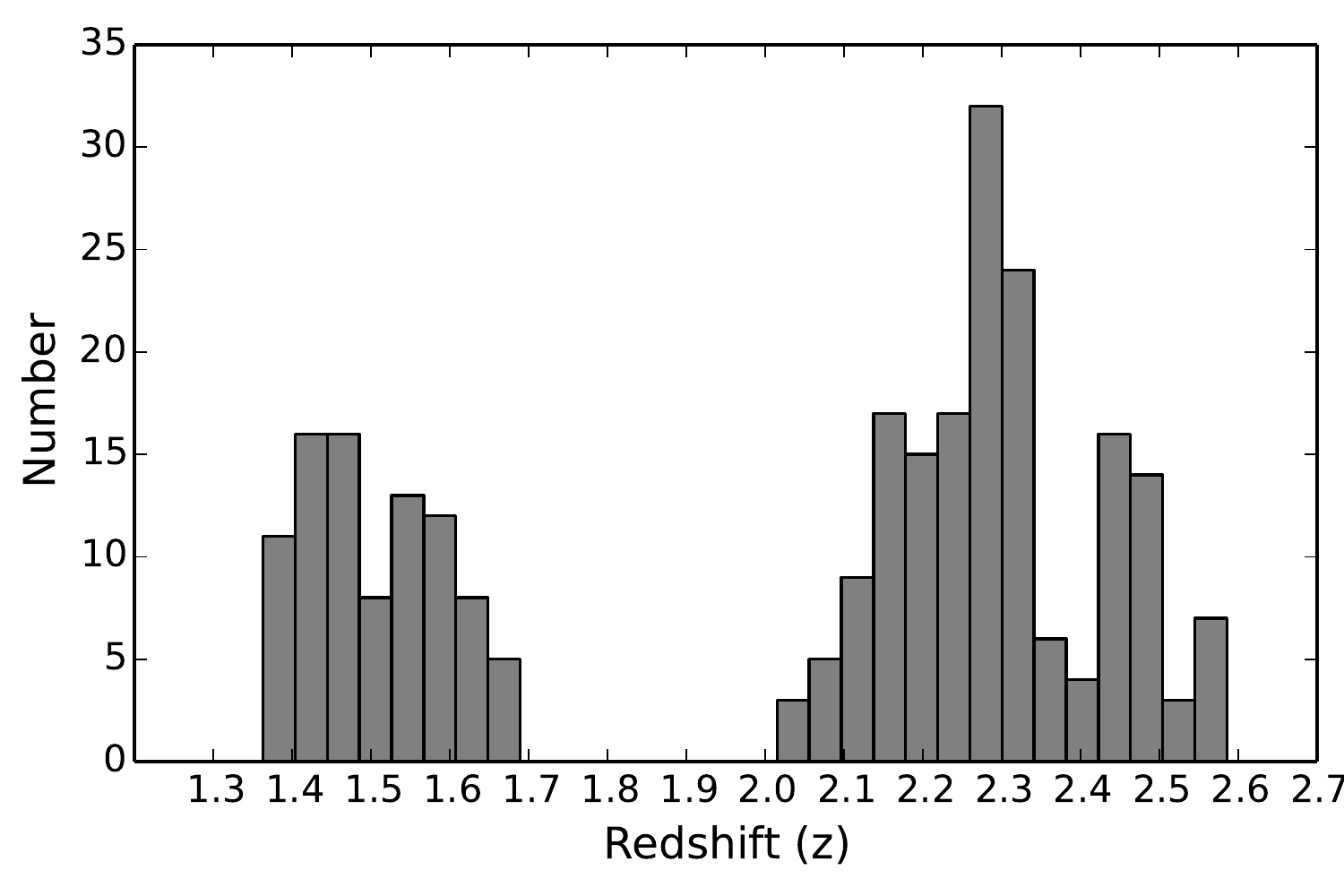}
		\caption{Redshift distribution of 261 star-forming galaxies in our sample with {\halpha} detections and coverage of the {\hbeta} line. There are 88 objects at $z\sim 1.5$ and 173 at $z\sim 2.3$. }
	\end{center}
	\label{fig:zhist}
\end{figure}

To assess the relation between SFR and $M_*$ for star-forming galaxies, we proceeded to remove all those galaxies that were deemed to be ``quiescent''.  In particular, we used a rest-frame {\em U}--{\em V} versus {\em V}--{\em J} color selection \citep{williams09} to identify the quiescent galaxies. There was only one object in the {\halpha}-{\hbeta} detected sample and two in the {\hbeta} undetected sample that were identified as quiescent and thus removed from the sample, leaving a final sample of 261 galaxies, with 215 detected in both lines. The redshift distribution is shown in Figure~\ref{fig:zhist}. 

\subsection{Line Flux and Balmer Absorption Measurements}

For each object, an initial redshift was estimated based on the emission line with the highest S/N ratio. The line fluxes were then measured by fitting Gaussian functions to the line profiles. The {\halpha} and [N{\sc ii}]$\lambda\lambda$6548,6584 doublet were fit using three Gaussian functions. To estimate the uncertainty in the line fluxes, the spectra were perturbed within the error spectra \footnote{Measuring the error spectra of the MOSDEF observations is described in detail in \citet{kriek15}. In brief, we constructed noise spectra by three independent methods: 1. based on the sky and readout noise level of the detector, 2. based on the variations between the science frames, and 3. based on the variations between the reduced spectra. The measured noise spectra based on the three methods agree with each other very well, showing that the noise measurements are robust. We used the first method error spectra in our analysis.} and all emission lines were re-fit. The dispersion among perturbed spectra was taken as the uncertainty in the line fluxes \citep{kriek15}.

An important aspect in determining the Balmer emission line fluxes is to correct for the underlying absorption that is produced in the atmosphere of primarily A-type stars. The Balmer absorption fluxes were determined from the best-fit SED models to the multi-wavelength photometry \citep{reddy15}, and were added to the emission line fluxes to get the total, absorption-corrected fluxes. For objects with {\halpha} ({\hbeta}) detected at $> 3\sigma$ level, ratio of the average {\halpha} ({\hbeta}) Balmer absorption flux to the average {\halpha} ({\hbeta}) observed flux is $\sim 2\% ~(\sim 15\%)$. The uncertainty in the absorbed fluxes is $\sim 2\%$ \citep{reddy15}, which is negligible compared to the typical uncertainty of our emission line fluxes $(\sim 15\%)$, and we decided not to include it in our flux uncertainty estimations.

\subsection{Mass and SFR Determination}
\label{sec:mass-sfr}
The stellar masses were determined through a $\chi^2$ minimization SED fitting procedure assuming \citet{bruzual03} models for solar metallicity, a Chabrier IMF, the \citet{calzetti00} dust attenuation curve with $0.0\leq E(B-V)\leq 0.6$, rising star-formation histories\footnote{ We should note that assuming a constant star-formation history for the SED modeling does not change the parameterization of the SFR-$M_*$ relation, as the scatter in the ratio of stellar masses inferred from a constant star-formation history modeling to the stellar masses derived assuming an exponentially rising star-formation history is $\sim 0.06$, which is smaller than the typical relative error of the stellar masses ($\sim 0.10$, see Section~\ref{sec:method}).} and a 50\,Myr lower limit on the age of the galaxies, assumed to be the typical dynamical time scale at this redshift \citep[see,][]{reddy15}. The SED-fitting was performed using photometry from the 3D-HST catalogs \citep{brammer12,skelton14} that has been corrected for the strongest emission lines measured in the MOSDEF survey, including [O{\sc ii}]$\lambda\lambda$3727,3730, [O{\sc iii}]$\lambda\lambda$4960,5008, {\hbeta}, and {\halpha} \citep{reddy15}. Errors in the SED parameters were estimated through Monte Carlo simulations; each photometric flux was randomly perturbed assuming a Gaussian distribution centered at the measured value with a standard deviation equal to the photometric error. This procedure was repeated many times and each photometric realization was refit to find the best-fit SED model. The standard deviation of the simulated SED parameters was taken as the error in the respective parameter.

We converted the {\halpha} luminosity to SFR using the \citet{kennicutt98} relation modified for a \citet{chabrier03} IMF, and corrected for dust attenuation using the nebular {\ebmv} derived from the Balmer decrement ({\halpha}/{\hbeta}), assuming a Galactic extinction curve taken from \citet*[][CCM]{cardelli89}. 
To determine the standard deviations in the corrected SFRs, we perturbed the measured {\halpha} and {\hbeta} fluxes according to their uncertainties by assuming a Gaussian distribution centered at the measured value with a standard deviation equal to the measurement error and recalculated the corrected SFRs based on each flux realization. We also accounted for spectroscopic slit losses\footnote{The slit loss uncertainty accounts for the loss of flux of a resolved source outside of the slit aperture. We calculated the slit loss uncertainty in our flux calibration by comparing the SED-inferred fluxes of the continuum detected galaxies in MOSDEF with their spectroscopic fluxes. For a full description, see Section 3.3, \citet{kriek15}.} by adding an additional uncertainty of 16\% (20\%) to the {\halpha} ({\hbeta}) measurement uncertainties in quadrature \citep{kriek15}.
Where {\hbeta} is undetected, a $3\sigma$ upper limit on the {\hbeta} flux was used to estimate a lower limit on the nebular {\ebmv} and the corresponding dust-corrected SFR.

We used the ancillary 3D-HST broad- and medium-band photometry at rest-wavelengths 1268 -- 2580~\AA~ to find the UV luminosity at 1600~\AA~ and the UV slope by fitting a power law to the photometry ($f_{\lambda}\propto \lambda^{\alpha}$). The UV slope was used to correct the UV luminosity for dust attenuation, following the \citet{meurer99} relation. We did not use the SED-inferred {\ebmv} because it is correlated with the stellar mass as both quantities are determined through SED fitting \footnote{As it can be seen in Table~\ref{tab:param}, the SFR(SED) and the SFR(UV) that is dust corrected by the {\ebmv} inferred from the SED modeling show tighter and steeper SFR-$M_*$ relations compared to that of SFR({\halpha}) and SFR(UV) corrected by the UV slope, indicating a stronger correlation with $M_*$. We should note that SFR(UV) corrected with the UV slope is still not completely uncorrelated with $M_*$ as the UV slope translates to an {\ebmv}.}. The luminosities were converted to SFRs using the \citet{kennicutt98} relation, adjusted for a Chabrier IMF.

\subsection{Stacking Procedure}
\label{sec:stack}
To further investigate the SFR-$M_*$ relation and the role of the {\hbeta}-undetected objects, we employed spectral stacks constructed in the following manner. Individual spectra were converted from flux density to luminosity density, normalized by the {\halpha} luminosity, and interpolated to a rest-frame grid with wavelength spacing of 0.5~\AA. 
The composite spectrum was then created by taking the average of the individual spectra, weighted by inverse variance of the error spectra. The composite error spectrum was estimated as the square root of the variance of the weighted mean.

The composite {\halpha} and {\hbeta} lines were fit with a Gaussian profile.
The errors on the line fluxes are derived using the standard deviation of the distribution of line fluxes by perturbing the stacked spectrum according to the composite error spectrum and recalculating the line fluxes 1000 times.
The composite line fluxes were corrected for Balmer absorption using the mean absorption, inversely weighted by flux error of the individual galaxies contributing to the bin (the individual Balmer absorptions were inferred from the best-fit stellar models).
The normalized composite spectrum was used to derive the {\halpha}/{\hbeta} ratio. To assess the total flux of the lines, we stacked the individual spectra without normalizing them by the individual {\halpha} luminosities.

\begin{figure}[tbp]
	\includegraphics[scale=.5]{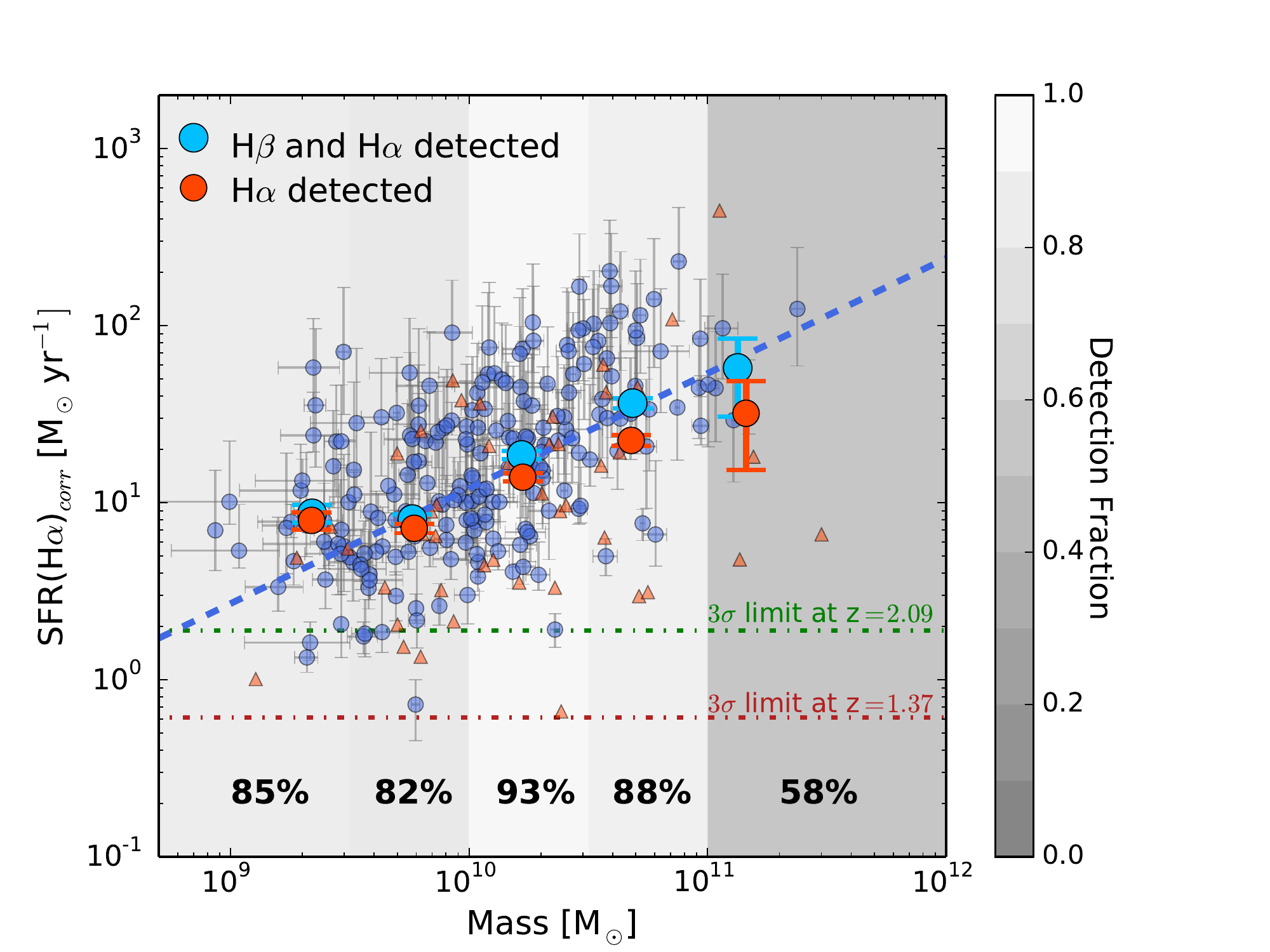}
	\caption{SFR({\halpha}) as a function of stellar mass for star-forming galaxies at $z$ = 1.37 -- 2.61, known as the star-forming main sequence. The individual points show galaxies with detected {\halpha} and {\hbeta} and the triangles are those with undetected {\hbeta}, for which $3\sigma$ lower limits of the dust corrected SFRs are plotted. The dashed line is the best linear fit to the individual detected galaxies with $\log(M_*/\msun) > 9.5$. The larger circles are stacks in five mass bins, taking into account only the detected galaxies (the light blue circles) and including the {\hbeta} non-detections (the red circles).
	The shaded regions and numbers on the bottom represent the fraction of galaxies with robust redshifts and {\halpha} detection with respect to the targeted objects in each mass bin.
	The $3\sigma$ limits on the SFR at $z=1.37$ and $z=2.09$ are calculated using the line sensitivities in {\em H} and {\em K} bands, respectively \citep{kriek15}, and are shown by dashed lines.
	}
	\label{fig:stack}
\end{figure}

\section{Analysis}
\label{sec:ms_analysis}
The correlation between stellar mass and SFR has been studied intensively since \citet{brinchmann04}. As a first step in studying this correlation in our {\halpha}- and {\hbeta}-detected sample, we calculated the Spearman’s rank correlation coefficient ($\rho$) between the log(SFR({\halpha})) and log($M_*$). We obtained $\rho\,=\,0.57$ at $8\,\sigma$, indicating a significant positive correlation between the two quantities. 

The SFRs versus stellar masses of individual galaxies are plotted in Figure~\ref{fig:stack}. The stacked SFRs in five stellar mass bins (the first bin is $\log(M_*/\msun)\,<\,9.5$ and the rest are divided by 0.5\,dex widths, Section~\ref{sec:stack}) are also plotted on top of the individual points. In this section, we describe our low-mass completeness, the fitting methodology, and discuss the effect of {\hbeta} nondetections and AGN on the SFR-$M_*$ relation.

\begin{figure}[tbp]
	\begin{center}
		\includegraphics[width=.5\textwidth]{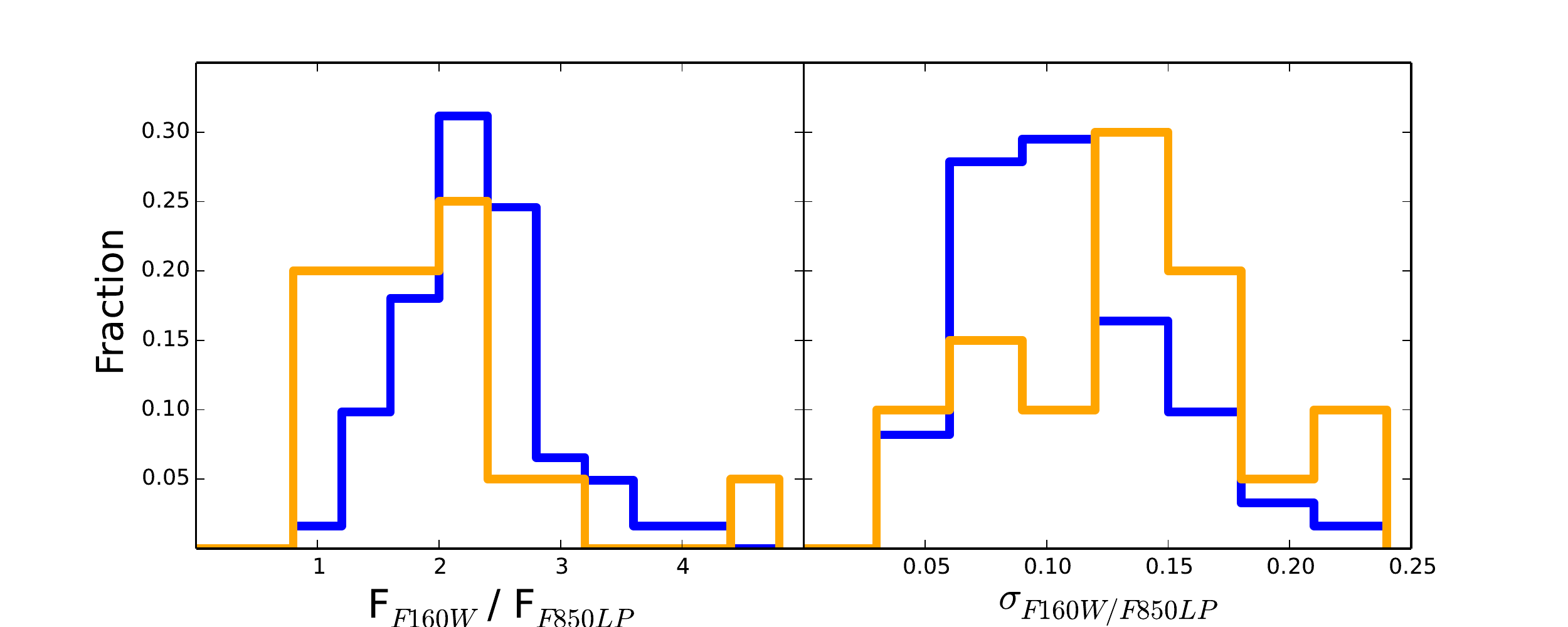} \quad
		\includegraphics[width=.5\textwidth]{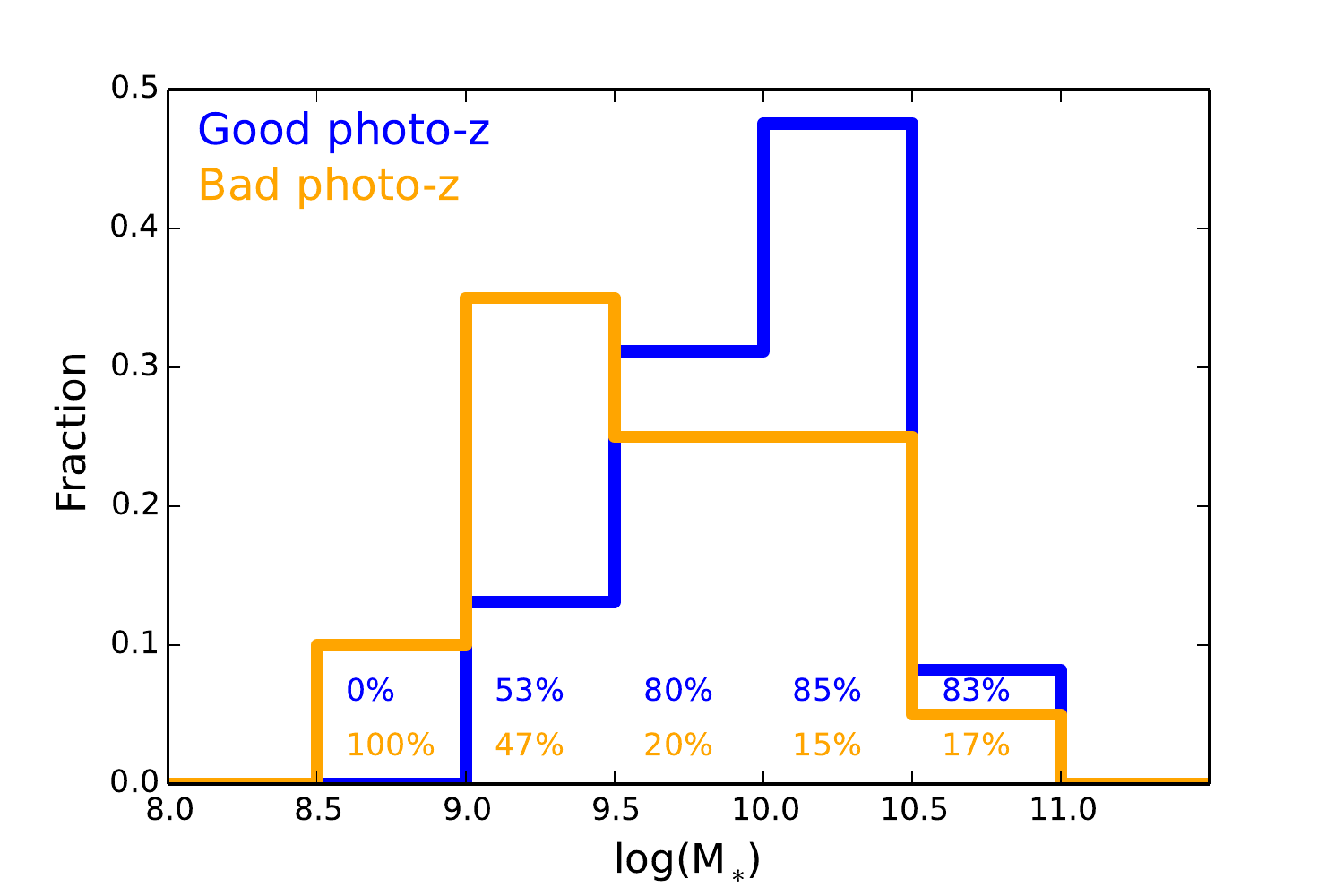}
		\caption{
		{\em Top:} Distribution of {\em F160W} to {\em F850LP} flux ratio (left) and the corresponding uncertainty (right) of UV-selected galaxies in the MOSDEF parent sample with spectroscopic redshifts at $z = 2.09-2.61$ \citep{reddy06b} in the GOODS-N field. The fluxes and uncertainties are from the 3D-HST photometric catalog \citep{skelton14}. The orange histogram denotes galaxies with photometric redshifts that fall outside of the spectroscopic redshift range (bad photo-z). The blue histogram shows galaxies with photometric redshifts in the same range as the spectroscopic redshift (good photo-z). The galaxies with bad photo-z's have on average bluer colors and larger uncertainties, indicative of a weaker Balmer/4000~\AA~ break compared to the good photo-z galaxies.
		{\em Bottom:} stellar mass distribution of the same samples described above. The orange/blue numbers indicate fraction of galaxies with bad/good photo-z's in each mass bin relative to the total number of galaxies in the corresponding bin.
		The histograms in these figures show the fractions, not the actual numbers. Therefore, the relative bin heights do not correspond to the percentage numbers on the bottom and should not be confused with each other.
		}
	\end{center}
	\label{fig:bias}
\end{figure}

\subsection{Mass Completeness}
\label{sec:mass_completeness}

As it is evident from Figure~\ref{fig:stack}, the stacked SFR of the first mass bin is elevated compared to the best-fit line. In this section, we discuss the completeness of our sample and address the possible causes of incompleteness at low stellar masses.

The redshifts of targets in the MOSDEF parent sample were adopted primarily from the 3D-HST grism measurements or other available spectroscopic campaigns \citep[see,][]{kriek15}. 
When a spectroscopic redshift was not available, which was the case for 67\% of the whole sample and 73\% of the objects with $M_*\,<\,10^{9.5}\msun$, redshifts were estimated using the 3D-HST photometry and the EAZY SED-fitting code \citep{brammer08}. 
Objects with secure photometric redshifts may be biased against those objects with weak Balmer/4000~\AA~ breaks (and hence younger ages and lower masses).  To assess the magnitude of this bias, we cross correlated our sample with the \citet{reddy06b} spectroscopic sample of UV-selected galaxies in the GOODS-N field. Rest-frame UV-selected samples are immune to the bias mentioned above as their selection criteria are not dependent on the strength of the Balmer or 4000~\AA~ break.

There were 81 galaxies at $z_{\rm spec} = 2.09-2.61$ that were in common between the MOSDEF and the UV-selected sample. We divided the common galaxies into two sets: those that have photometric redshifts in the same range as their spectroscopic redshifts of $z = 2.09-2.61$ (good photo-z, 61 objects), and those with photometric redshifts outside of this range (bad photo-z, 20 objects). The bad photo-z objects are not scattered symmetrically around $z_{\rm spec} = 2.09-2.61$; their photometric redshifts are systematically lower than our target redshift range, i.e., $z_{\rm phot}\,<\,2.09$.

The two filters that bracket the Balmer break are {\em HST}/WFC3 {\em F160W} and {\em HST}/ACS {\em F850LP}, which were available for all 81 galaxies examined here. As expected, the galaxies with incorrect photometric redshifts had on average bluer {\em F850LP} - {\em F160W} colors, with larger uncertainties. The distributions of the flux ratios and the associated uncertainties are plotted in Figure~\ref{fig:bias}.
Were it not for prior spectroscopic redshifts, those galaxies with bad photo-z's would have been missing from our sample. The conclusion is further supported by the two-sided Kolmogorov-Smirnov test, which indicates a low probability of $P\simeq 0.017$ that the {\em F850LP} - {\em F160W} color distribution of the good and bad photo-z sets are drawn from the same parent distribution.

The stellar-mass distributions of the two sets of galaxies are shown in the bottom panel of Figure~\ref{fig:bias}, with the fraction of galaxies in each mass bin written on the bottom of the plot. The distributions indicate that the majority of galaxies with stellar masses below $10^{9.5}\msun$ would have been missed from the MOSDEF sample if we had relied on their photometric redshifts. Therefore, to avoid any biases we limited our analysis to those with $M_* > 10^{9.5}\msun$. There are 128 and 185 galaxies above $10^{9.5}\msun$ at $z = 2.09-2.61$ and $z = 1.37-2.61$, respectively.
The dashed line in Figure~\ref{fig:stack} is the linear regression fit to the individual galaxies with $M_* > 10^{9.5}\msun$. 
The low mass limit we adopted is also the limit above which the inferred log(SFR)-log($M_*$) slope (see Section~\ref{sec:method}) deviates by greater than $1\sigma$ from the slope over the full stellar mass range, including the low mass galaxies.

The other possible bias at the low-mass end of our sample is a Malmquist bias, due to the fact that our requirement of an {\halpha} detection may bias against objects with faint {\halpha} emission. To test this possibility, we calculated the fraction of galaxies that were detected in {\halpha} to the number of galaxies targeted for spectroscopy, in bins of stellar mass. The fractions are listed on the bottom of Figure~\ref{fig:stack}.
The detection fraction of the first bin (the lowest mass bin) is similar to that of the higher mass bins, indicating that by limiting our sample to the {\halpha}-detected objects we did not miss a significantly higher fraction of galaxies at low masses.

The rest-frame optically selected sample of the MOSDEF survey is less prone to missing highly obscured galaxies at high stellar masses, compared to the UV-selected samples. The effect of this potential bias of UV-selected samples versus our sample will be discussed in Section~\ref{sec:slope_sample}.

\capstartfalse   
\begin{deluxetable*}{cccccc}
\setlength{\tabcolsep}{0.05in} 
\tabletypesize{\footnotesize} 
\tablewidth{0pc}
\tablecaption{Parameters of the $\log(\text{SFR})-\log(M_*)$ Linear Fit \tablenotemark{a}}
\tablehead{
\colhead{Redshift Range} &
\colhead{SFR Indicator} &
\colhead{Slope} &
\colhead{Intercept} &
\colhead{Observed Scatter} &
\colhead{Measurement-Subtracted Scatter}
}
\startdata
{$z = 1.37 - 2.61$} & {\halpha}\tablenotemark{c} & $0.65\,\pm\,0.08$ & $-5.40\,\pm\,0.86$& $0.40$ & $0.36$ \\
{(N = 185)\tablenotemark{b}} & UV$_{\beta}$\tablenotemark{d} & $0.62\,\pm\,0.08$ & $ -5.03\,\pm\,0.80$ & $0.34$ & $0.30$ \\
{} & SED\tablenotemark{e} & $0.80\,\pm\,0.05$& $-6.79\,\pm\,0.55$ & $0.28$ & $0.25$ \\
{} & UV$_{\text{SED}}$\tablenotemark{f} & $0.79\,\pm\,0.07$& $-6.58\,\pm\,0.67$ & $0.27$ & $0.20$ \\
{} & {} & {} & {} & {} & {} \\
{$z = 2.09 - 2.61$} & {\halpha}\tablenotemark{c} & $0.58\,\pm\,0.10$ & $-4.65\,\pm\,1.05$& $0.36$ & $0.31$ \\
{(N = 128)\tablenotemark{b}} & UV$_{\beta}$\tablenotemark{d} & $0.71\,\pm\,0.09$ & $ -5.84\,\pm\,0.92$ & $0.31$ & $0.25$ \\
{} & SED\tablenotemark{e} & $0.83\,\pm\,0.07$& $-7.02\,\pm\,0.73$ & $0.29$ & $0.25$ \\
{} & UV$_{\text{SED}}$\tablenotemark{f} & $0.75\,\pm\,0.09$& $-6.11\,\pm\,0.95$ & $0.28$ & $0.18$
\enddata
\tablenotetext{a}{The parameters are calculated for galaxies with $M_* > 10^{9.5}\msun$, using an OLS linear regression method}
\tablenotetext{b}{Number of objects}
\tablenotetext{c}{The {\halpha} SFR is dust corrected by Balmer decrement, assuming the Cardelli Galactic extinction curve}
\tablenotetext{d}{The UV SFR is dust corrected by the UV slope, assuming the Calzetti attenuation curve}
\tablenotetext{e}{The SFR inferred from the stellar population model, assuming the Calzetti curve for dust attenuation}
\tablenotetext{f}{The UV SFR is dust corrected by the {\ebmv} inferred from the SED model, assuming the Calzetti attenuation curve}
\label{tab:param}
\end{deluxetable*}
\capstarttrue  

\subsection{Methodology}
\label{sec:method}

There are various linear regression methods that can be used to characterize a correlation between two variables. 
In most studies of the SFR-$M_*$ relation, the correlation between SFR and $M_*$ has been determined using the ordinary least-squares regression (OLS). In the OLS method, one variable (here, the SFR) is assumed as the dependent variable and the regression line is defined to be that which minimizes the sum of the squares of the vertical distances between the data points and the regression line. The OLS method should be used in science cases where there is an independent variable that is estimated confidently and will be used to predict the other variable. 
As over a galaxy's lifetime the stellar mass is built relatively smoothly as opposed to the SFR, which can be stochastic and vary to a greater extent, we are more interested in determining the scatter and mean of SFR as a function of stellar mass. For this purpose, the OLS method is a valid statistical procedure to be used for characterizing the SFR-$M_*$ relation.

We determined the best estimates for the OLS regression slope and intercept by perturbing the data according to the stellar mass and SFR uncertainties 1000 times. The median of the slope and intercept distributions are taken as the best-fit parameters.
The slope and intercept uncertainties were calculated based on the OLS method error estimates in \citet{isobe90}. \citet{isobe90} estimated the errors of regression parameters for linear relations where there is an intrinsic scatter. The average of the uncertainties obtained for the individual realizations was adopted as the uncertainty in the regression coefficients. We found a slope of $0.65\,\pm\,0.08$ for SFR({\halpha}) at 1.37$\,<\,z\,<\,$2.61, and $0.58\,\pm\,0.10$ for the 2.09$\,<\,z\,<\,$2.61 sample only, with a lower mass limit of $10^{9.5}\msun$. The relation is presented in Figure~\ref{fig:stack} and Table~\ref{tab:param}. The SFR-$M_*$ relation was not investigated separately for the sample at $1.37\,<\,z\,<\,1.70$ due to the small number of galaxies at this redshift range.

In the OLS method, the scatter is defined as the scatter in SFR at a certain stellar mass.
We estimated the total observed scatter by calculating the 3-$\sigma$ clipped dispersion in the distribution of vertical distances from the best-fit line. We propagated the mean measurement errors of mass and SFR\footnote{The uncertainty due to spectroscopic slit losses is included in the SFR error estimates as well (Section~\ref{sec:mass-sfr}).} and subtracted them (in quadrature) from the total observed scatter to obtain the measurement-subtracted scatter. We found a measurement-subtracted scatter of 0.31~dex for SFR({\halpha}) at $2.1\leq z\leq 2.6$. The measurement-subtracted scatter still includes galaxy-to-galaxy variations in the assumed dust attenuation curve (Section~\ref{sec:scatter}).

An alternative statistical procedure to the OLS method is a regression method that treats the two variables symmetrically, such as the OLS bisector method. The OLS bisector is recommended to be applied in cases where an underlying functional relation between two independent variables is studied \citep{isobe90}. The scatter in this method is the scatter about the best-fit line to the relation. For comparison, we reported the OLS bisector values in Table~\ref{tab:param_bisec}. We should emphasize that as we are most interested in determining the scatter and mean of SFR at a given stellar mass, the OLS method is the one on which we based our main conclusions.

\capstartfalse   
\begin{deluxetable*}{cccccc}
\setlength{\tabcolsep}{0.05in} 
\tabletypesize{\footnotesize} 
\tablewidth{0pc}
\tablecaption{Parameters of the $\log(\text{SFR})-\log(M_*)$ Linear Fit, Using The OLS Bisector Method \tablenotemark{a}}
\tablehead{
\colhead{Redshift Range} &
\colhead{SFR Indicator} &
\colhead{Slope} &
\colhead{Intercept} &
\colhead{Observed Scatter} &
\colhead{Measurement-Subtracted Scatter}
}
\startdata
{$z = 1.37 - 2.61$} & {\halpha}\tablenotemark{c} & $1.26\,\pm\,0.07$ & $-11.57\,\pm\,0.69$ & 0.33 & 0.30 \\
{(N = 185)\tablenotemark{b}} & UV$_{\beta}$\tablenotemark{d} & $1.14\,\pm\,0.06$ & $ -10.33\,\pm\,0.62$ & 0.29 & 0.26
\enddata
\tablenotetext{a}{The parameters are calculated for galaxies with $M_* > 10^{9.5}\msun$}
\tablenotetext{b}{Number of objects with $M_* > 10^{9.5}\msun$}
\tablenotetext{c}{The {\halpha} SFR is dust corrected by Balmer decrement, assuming the Cardelli Galactic extinction curve}
\tablenotetext{d}{The UV SFR is dust corrected by the UV slope, assuming the Calzetti attenuation curve}
\label{tab:param_bisec}
\end{deluxetable*}
\capstarttrue  

\begin{figure*}[tbp]
\includegraphics[width=.5\textwidth]{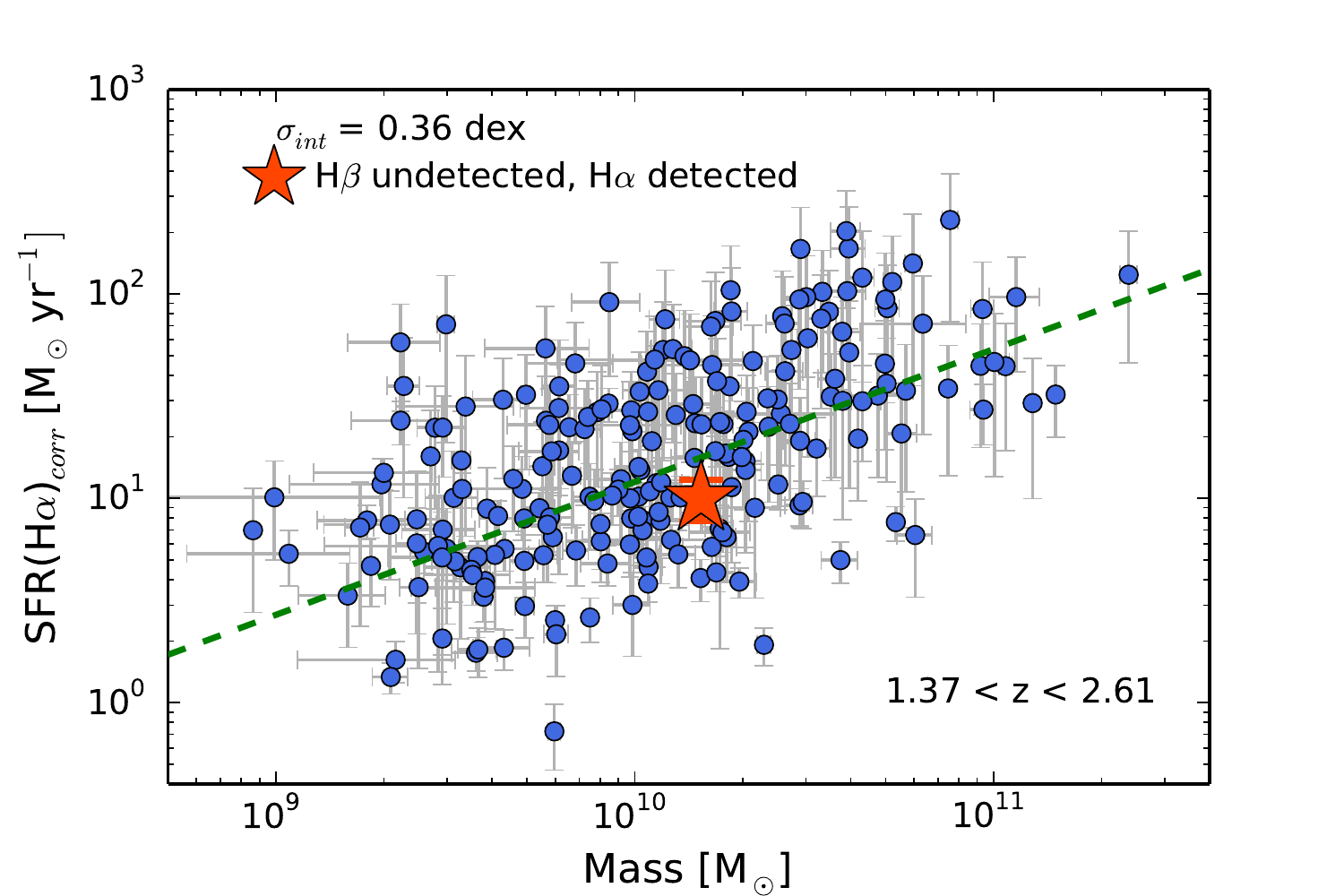}\quad
\includegraphics[width=.5\textwidth]{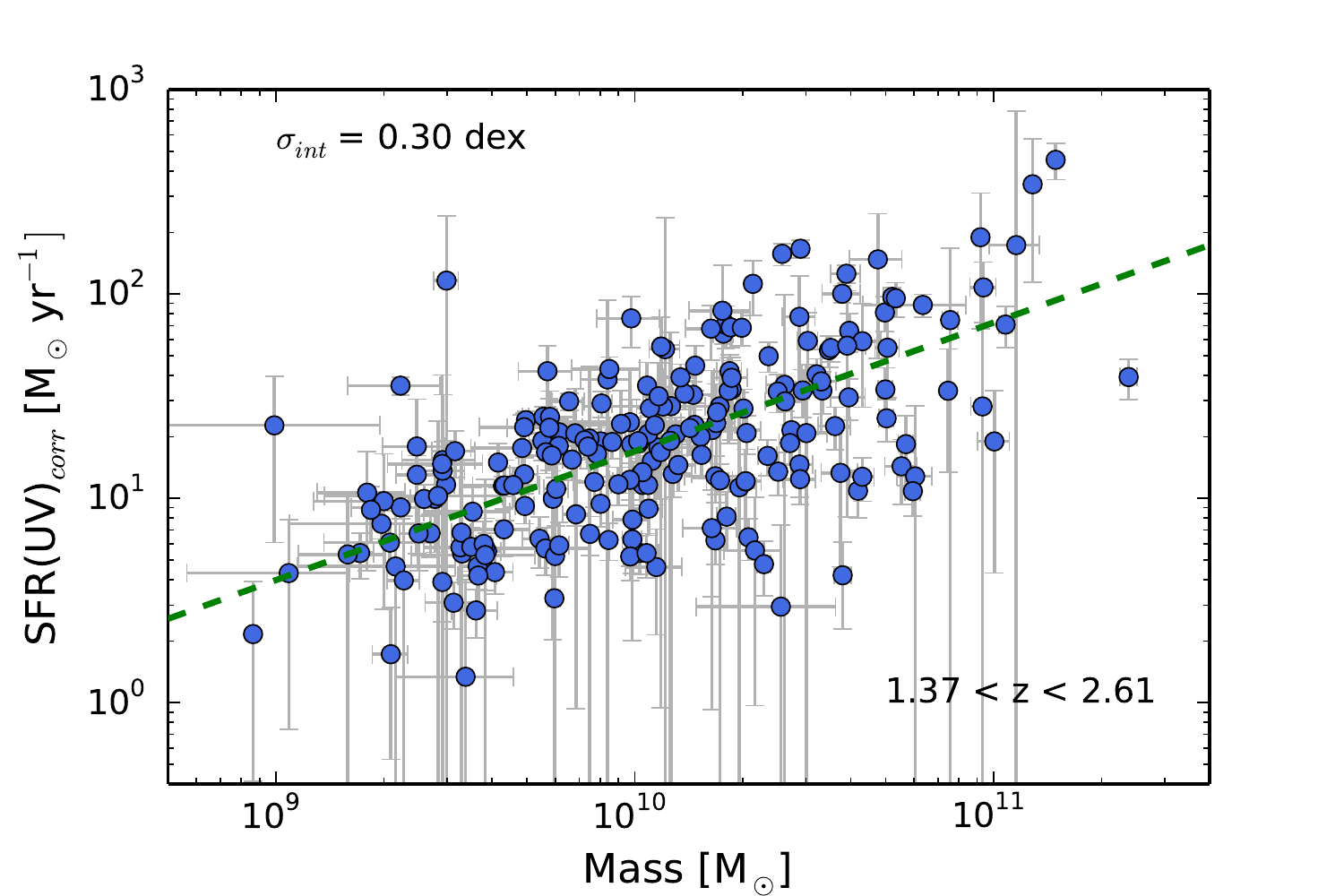}
	\caption{SFR as a function of stellar mass for star-forming galaxies at $z = 1.37 - 2.61$. {\em Left:} SFR({\halpha}) -- corrected for dust attenuation assuming the Balmer decrement and a \citet{cardelli89} extinction curve -- versus $M_*$. The red star shows the weighted-average stack of the {\hbeta}-undetected galaxies, where its mass is the median of the mass of galaxies in the stack. The stacked point is within the SFR({\halpha}) scatter of 0.36\,dex, suggesting that those galaxies undetected in {\hbeta} have on average a similar proportion of SFR given their stellar mass, as those galaxies with both lines detected.
	{\em Right:} SFR(UV) versus $M_*$. The SFR is inferred from the UV luminosity at 1600\AA~ and is dust corrected by the UV slope, both determined from the multi-band photometry at 1268--2580\AA~\citep{skelton14}. The green lines show the regression lines fitted to $\log(\text{SFR})-\log(M_*)$ relation. The measurement-subtracted scatter in SFR (i.e., the scatter after the subtraction of SFR and mass measurement uncertainties) is provided on the upper left corners of each plot.
	}
	\label{fig:scatter}
\end{figure*}

\subsection{Nondetections}
It is important to investigate whether excluding the {\hbeta} nondetections from the main analysis causes any biases. In other words, do the {\hbeta}-undetected galaxies represent different physical characteristics or are they undetected due to observational shortcomings? 
We stacked the spectra of the {\hbeta}-undetected objects as described in Section~\ref{sec:stack}, and derived the dust-corrected SFR({\halpha}) using the stacked spectrum. In the composite spectrum, the stacked {\hbeta} line is detected at $3.9\sigma$. The stacked SFR and median stellar mass of the undetected-{\hbeta} objects are plotted on top of the SFR-$M_*$ relation in Figure~\ref{fig:scatter} as the red star. 
The star lies below the linear fit by 2.5$\sigma$, but is still within the SFR({\halpha}) scatter of 0.36\,dex, implying that these objects are not on average intrinsically different from the detected objects and excluding them should not systematically bias the SFR-$M_*$ relation. As mentioned in Section~\ref{sec:hasample}, this conclusion is further supported by the fact that the spectra of a large fraction of these galaxies are contaminated by skylines.

\subsection{AGN}
As mentioned before in Section~\ref{sec:sample}, AGN were identified through X-ray, IR, and optical line observations and removed from the sample. The AGN mostly reside on the high mass end of the main sequence and it is interesting to examine how they would change the slope and scatter of the log(SFR)-log($M_*$) relation. Here, we assumed that all of the rest-frame UV continuum emission in AGN host galaxies is coming from the stellar population; we determined the SFRs from the UV luminosity at 1600 \AA, dust corrected using the UV slope, and estimated the masses through the SED modeling. Including the AGN does not change the slope significantly ($0.66\,\pm\,0.09$) but results in a larger scatter (0.41\,dex) that is mainly occurring at the massive end, as most of the AGN have larger stellar masses than the typical stellar mass of the star-forming objects. This is a well-understood selection effect for AGN that they are easier to detect in more massive galaxies \citep{aird12}.

\section{The Scatter of The log(SFR)-log($M_*$) Relation}
\label{sec:scatter}

In this section, we investigate the intrinsic scatter in the log(SFR)-log($M_*$) relation by using different SFR indicators and considering variations in the dust attenuation curve.
The observed and measurement-subtracted scatter in SFR({\halpha}) versus $M_*$ and SFR(UV) versus $M_*$, calculated as described in Section~\ref{sec:method}, are presented in Table~\ref{tab:param} and Figure~\ref{fig:scatter}. For a fair comparison, we limited our UV sample to the same sample as the {\halpha} detected galaxies. 

The measurement-subtracted scatter reported here is affected by uncertainties such as galaxy-to-galaxy variations in IMF and dust attenuation curve. In order to test the effect of dust attenuation variations, we simulated an SFR-$M_*$ relation with 1000 galaxies and an intrinsic scatter of 0.2\,dex, assuming the log(SFR)-log($M_*$) slope and intercept derived based on our MOSDEF data (Section~\ref{sec:method}). We attenuated the intrinsic SFRs to derive the observed (unobscured) SFRs following the relation, log(SFR$_{\text{obs}}$)$=$log(SFR$_{\text{int}})\,-\,0.4\,\kappa_{\lambda}$\,{\ebmv}. In this equation, the {\ebmv} was calculated using a linear correlation between the nebular {\ebmv} and the intrinsic SFR with a scatter of 0.5\,dex, based on the MOSDEF data. The reddening value, $\kappa_{\lambda}$, was selected randomly according to a valid range of the most commonly used attenuation curves in the literature; $\kappa_{\halpha}$ varied between the MOSDEF \citep{reddy15} and the \citet{calzetti00} curves' reddening values, and $\kappa_{1600}$ varied between the Milky Way and the SMC \citep{gordon03} reddening values. The dust-corrected SFRs were then reproduced by assuming a {\em single} $\kappa_{\lambda}$ (the Calzetti value for the SFR(UV) and the Cardelli value for the SFR({\halpha})). The scatter in the new dust-corrected log(SFR({\halpha}))-log($M_*$) relation was $\sim\,0.02$\,dex larger than the intrinsic simulated scatter, while the dust-corrected log(SFR(UV))-log($M_*$) scatter increased by $\sim\,0.10$\,dex. The larger increase in the SFR(UV) scatter is followed by larger variations of the attenuation curves at shorter UV wavelengths compared to the variations of the attenuation curves at optical wavelengths\footnote{Here, we assumed that the scatter in $\kappa_{\lambda}$ does not change as a function of luminosity or stellar mass, but we know that the scatter in $\kappa_{1600}$ is larger at higher UV luminosities \citep[e.g.,][]{reddy10,reddy15}. This effect would increase the recovered scatter in log(SFR)-log($M_*$) relation even to a greater degree.}. This simple simulation indicates that without knowing the attenuation curve variations, it is not trivial to interpret the difference in the SFR({\halpha}) and the SFR(UV) scatter solely as the SFR stochasticity due to different timescales.
Variations in the assumed IMF and metallicity would also affect the SFR({\halpha}) and SFR(UV) scatter to different degrees, again making it difficult to conclude about the burstiness of the star formation based on the difference in the scatter of SFR({\halpha}) and SFR(UV). It is worth pointing out that picking a random star-formation history between a constant and an exponentially rising star-formation history affects the SED inferred stellar masses by $\lesssim 0.04$\,dex, which is negligible compared to the scatter in the log(SFR)-log($M_*$) relation.

The calculated measurement-subtracted scatter in the log(SFR({\halpha}))-log($M_*$) relation is 0.06\,dex larger than that of the log(SFR(UV))-log($M_*$) relation. As the two SFR indicators trace SFR on different timescales\footnote{The {\halpha} luminosity traces SFR on shorter timescales compared to the UV luminosity, as {\halpha} is sensitive to the most massive ($M_* \ga 15\msun$) and short-lived (ages $\la 10$\,Myr) stars, while UV traces stars with $M_* \ga 5\msun$ and ages $\la 100$\,Myr, \citep{kennicutt98,kennicutt12,madau14}.}, the difference in SFR scatter may be attributed to the stochasticity of SFR.
It has been suggested in the galaxy formation simulations that for a massive galaxy ($z=0$ halo mass of $10^{12}\msun$) with a typical merger history, the dispersion in SFR averaged over timescales of $\sim 10^8$ years could be $\sim 0.03 - 0.1$\,dex smaller than the scatter of SFR averaged over shorter timescales of $\sim 10^7$ years \citep{hopkins14,sparre15,dominguez15}. 
On the other hand, in our study the {\halpha} light is captured through a slit while the UV light is obtained from imaging data, which may introduce additional scatter to our measurements. Considering these uncertainties and poorly constrained effects of the variations in dust attenuation curve and IMF, as discussed earlier in this section, we conclude that the 0.06\,dex difference in scatter is not significant and can not be attributed solely to the SFR stochasticity in these galaxies.

According to \citet{speagle14}, the intercept of the log(SFR)-log($M_*$) relation changes as $0.12\,\pm\,0.04 \times t$, where $t$ is the age of the universe in Gyr. This change corresponds to an offset of 0.09\,dex from $z=2.0$ to 2.6. The offset has a negligible effect on the 0.31\,dex scatter of log(SFR({\halpha}))-log($M_*$) relation derived here at $2.0<\,z<\,2.6$, and subtracting it in quadrature will reduce the scatter to 0.30\,dex.

Prior to the use of multi-wavelength estimates of the SFR (e.g., SFR(IR)+SFR(UV)), it was common to assess the SFR-$M_*$ relation using SED-inferred SFRs \citep[e.g.,][]{magdis10,sawicki12}. To investigate how SFR(SED) would affect the relation, we also calculated the scatter using SED-inferred SFR and UV SFR, where the latter was corrected by the SED-derived {\ebmv} rather than the UV slope. 
The scatter in SFR is reduced to 0.25 and 0.20\,dex for SFR(SED) and SFR(UV), respectively (the last two rows in Table~\ref{tab:param}).
These values clearly underestimate the true scatter in the log(SFR)-log($M_*$) relation, because SFR(SED) and SED-inferred {\ebmv} are highly correlated with the SED-derived stellar masses.

Galaxy formation simulations suggest various drivers for the scatter in log(SFR)-log($M_*$) relation. \citet{dutton10a} argued that the source of the small scatter they derived (0.12\,dex) is the variance in mass accretion histories for halos of a given virial mass, but \citet{dave12} suggested that the more complex relation between the gas inflow rate and the dilution time, which shows how quickly galaxies can return to equilibrium after being perturbed away from the main sequence also influences the scatter. 
Our results show that the larger scatter we found ($\sim 0.30-0.40$\,dex) is not only due to the measurement uncertainties, which indicates that the gas accretion and star-formation histories are not as smooth as predicted by the some simulations. Also, by assuming an average attenuation curve and a single IMF and metallicity for all the galaxies contributing to the SFR-$M_*$ relation, the measurement-subtracted scatter is a lower limit on the intrinsic scatter in SFRs at a given stellar mass.

\section{The Slope of The log(SFR)-log($M_*$) Relation }
\label{sec:slope}

In this section, we focus on the slope of the log(SFR)-log($M_*$) relation for the SFR({\halpha}) sample.
The inferred slope (see Section~\ref{sec:method} for the methodology) is $0.58\,\pm\,0.10$ for the $z\sim 2.3$ sample and when we include the $z\sim 1.5$ sample the slope is $0.65\,\pm\,0.08$, at $M_* > 10^{9.5}\msun$ (refer to Section~\ref{sec:mass_completeness} for the discussion on the low-mass limit). The slope of the full sample (i.e., $M_* > 10^{9.0}\msun$) is $0.56\,\pm\,0.09$.

There are many studies in the literature that parameterize the SFR-$M_*$ relation and their reported slopes span in a wide range of $\sim 0.3 - 1.0$ \citep[ e.g.,][]{daddi07a,dunne09,pannella09,santini09,rodighiero11,karim11,zahid12,reddy12b,whitaker12b,sobral14,atek14,rodighiero14}. Various effects could lead to discrepancies in the slope estimates. In this section, we try to address three of the most significant effects by investigating them in our sample.

\begin{figure}[tbp]
	\includegraphics[scale=.55]{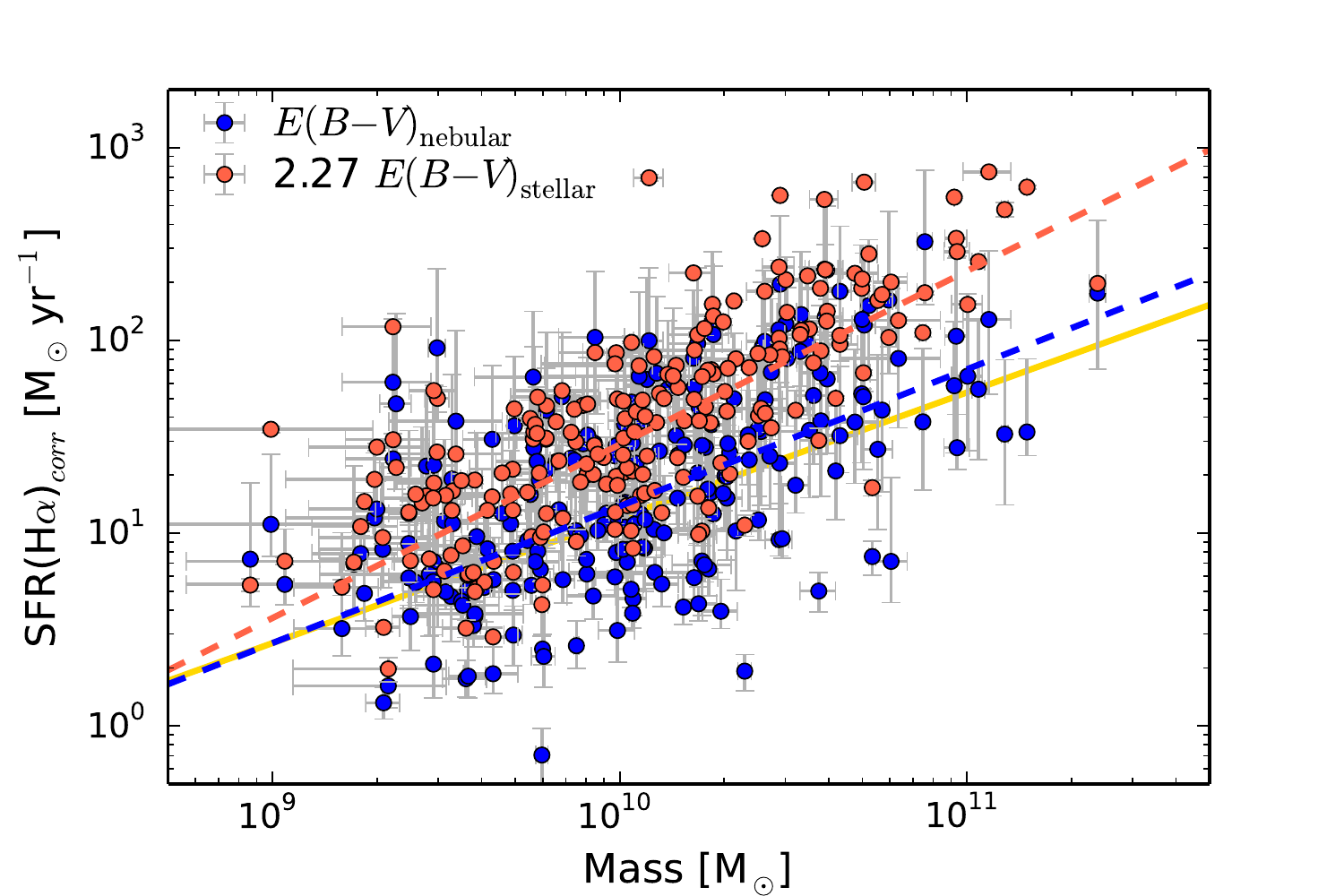} \quad
	\includegraphics[scale=.55]{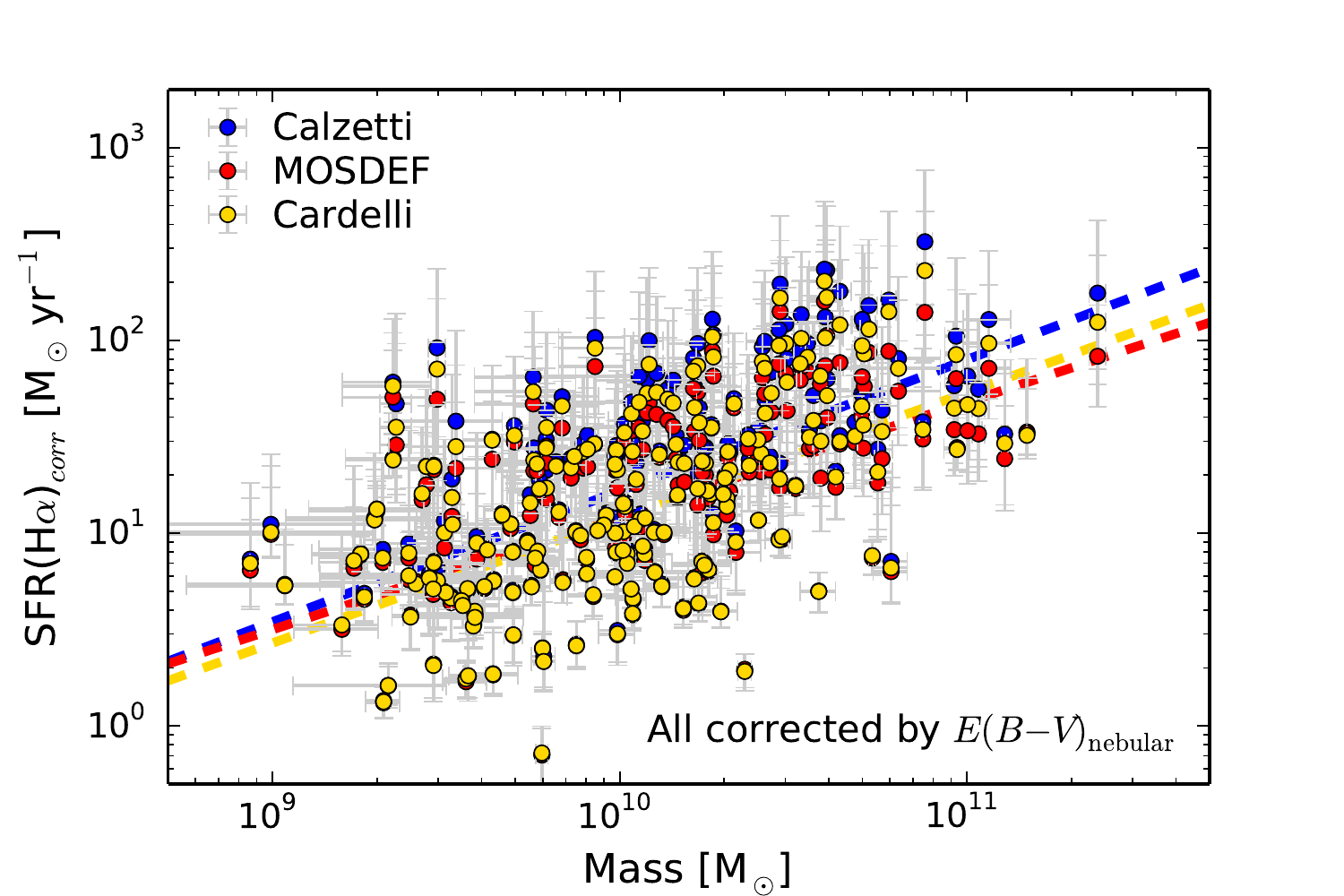}
	\caption{The SFR({\halpha})-$M_*$ relation, where the SFR is corrected for dust attenuation by different methods and attenuation curves. {\em Top:} The blue dots show SFR({\halpha}) dust corrected by the nebular {\ebmv} (derived from the observed Balmer decrement) and assuming a Calzetti curve. The red dots represent the SFR({\halpha}) corrected by the SED inferred {\ebmv} multiplied by 2.27, again assuming a Calzetti curve. The latter is a method that is commonly used in high-redshift studies to correct the {\halpha} luminosities                                                                                                                without having access to the Balmer decrement. This method overestimates the corrected SFRs and results in a steeper slope ($0.90\,\pm\,0.07$) compared to the nebular corrected SFRs (slope = $0.71\,\pm\,0.09$). The dashed lines show the linear fits to galaxies with $\log(M_*/\msun) > 9.5$. The yellow line is the linear fit to galaxies corrected by nebular {\ebmv} but assuming a Cardelli Galactic extinction curve, which we used as a default in this study.
	{\em Bottom:} The SFR-$M_*$ relation for the same set of galaxies, corrected with the nebular {\ebmv} assuming three different dust curves: the Calzetti curve, the MOSDEF curve, and the Cardelli curve. The best-fit lines are shown with the relevant colors. All three slopes are consistent with each other within their uncertainties.
	}
	\label{fig:sfrha_dust}
\end{figure}

\subsection{Dust Correction}
The method used for dust correction affects the derived slope of the log(SFR)-log($M_*$) relation. In high-redshift studies, a commonly used dust correction method for the nebular lines is to multiply the {\ebmv} derived for the stellar continuum by 2.27 factor and assume the \citet{calzetti00} attenuation curve. It has been shown in some previous studies that this recipe overestimates the corrected {\halpha} SFRs for UV-selected galaxies \citep[e.g.,][]{reddy10,steidel14,shivaei15}. 
In the top panel of Figure~\ref{fig:sfrha_dust} we compared the {\halpha} SFRs corrected by $E(B-V)_{\rm nebular}$, computed from the Balmer decrement, with those corrected by $2.27\times E(B-V)_{\rm SED}$. We emphasize that this practice was done solely for the purpose of demonstrating how the commonly used dust correction recipes could potentially affect the parameterization of the SFR-$M_*$ relation in the absence of Balmer decrement measurements. The SFRs corrected with $2.27\times E(B-V)_{\rm SED}$ assuming the Calzetti curve yield a main sequence slope of $0.90\,\pm\,0.05$, while the slope of those corrected by the $E(B-V)_{\rm nebular}$ with the Calzetti curve is $0.71\,\pm\,0.09$. The steeper slope is expected as the color-excess ({\ebmv}) increases with stellar mass \citep[e.g.,][]{garnbest10} and multiplying {\ebmv} by a factor (here, 2.27) causes a more significant difference at higher masses. Adopting a value between 1 and 2.27 \citep{kashino13,wuyts13} will make the slope shallower than 0.90 but still steeper than the 0.71 slope derived by the $E(B-V)_{\rm nebular}$ with the Calzetti curve.

We also investigated the potential effect of the assumed attenuation curve on the best-fit slope of the log(SFR)-log($M_*$) relation.
We corrected the observed {\halpha} SFRs assuming different attenuation curves: the Calzetti curve \citep{calzetti94}, the attenuation curve derived from galaxies in the MOSDEF sample \citep{reddy15}, and the CCM curve \citep{cardelli89}. The first two are stellar attenuation curves while the latter is a line of sight extinction curve that we used to correct the {\halpha} luminosities throughout this paper.
The results are presented in the bottom panel of Figure~\ref{fig:sfrha_dust}. The nebular color excess (i.e., $E(B-V)_{\rm nebular}$) does not change significantly between the three curves, the reason being that the shapes of these three curves in the rest-frame optical are very similar. The main difference arises from normalization of the curves, affecting the total attenuation correction. \citet{reddy15} showed that assuming the Calzetti curve for the nebular lines as opposed to the CCM curve results in higher SFRs, where the difference increases with increasing SFR (see Figure~21 in \citealt{reddy15}). We see the same trend here in Figure~\ref{fig:sfrha_dust}, namely that the discrepancy between corrected SFRs is larger at higher SFRs. Despite the change in SFR, the three main sequence slope estimates are consistent with each other within their uncertainties.

\begin{figure}[tbp]
	\begin{center}
		\includegraphics[scale=.55]{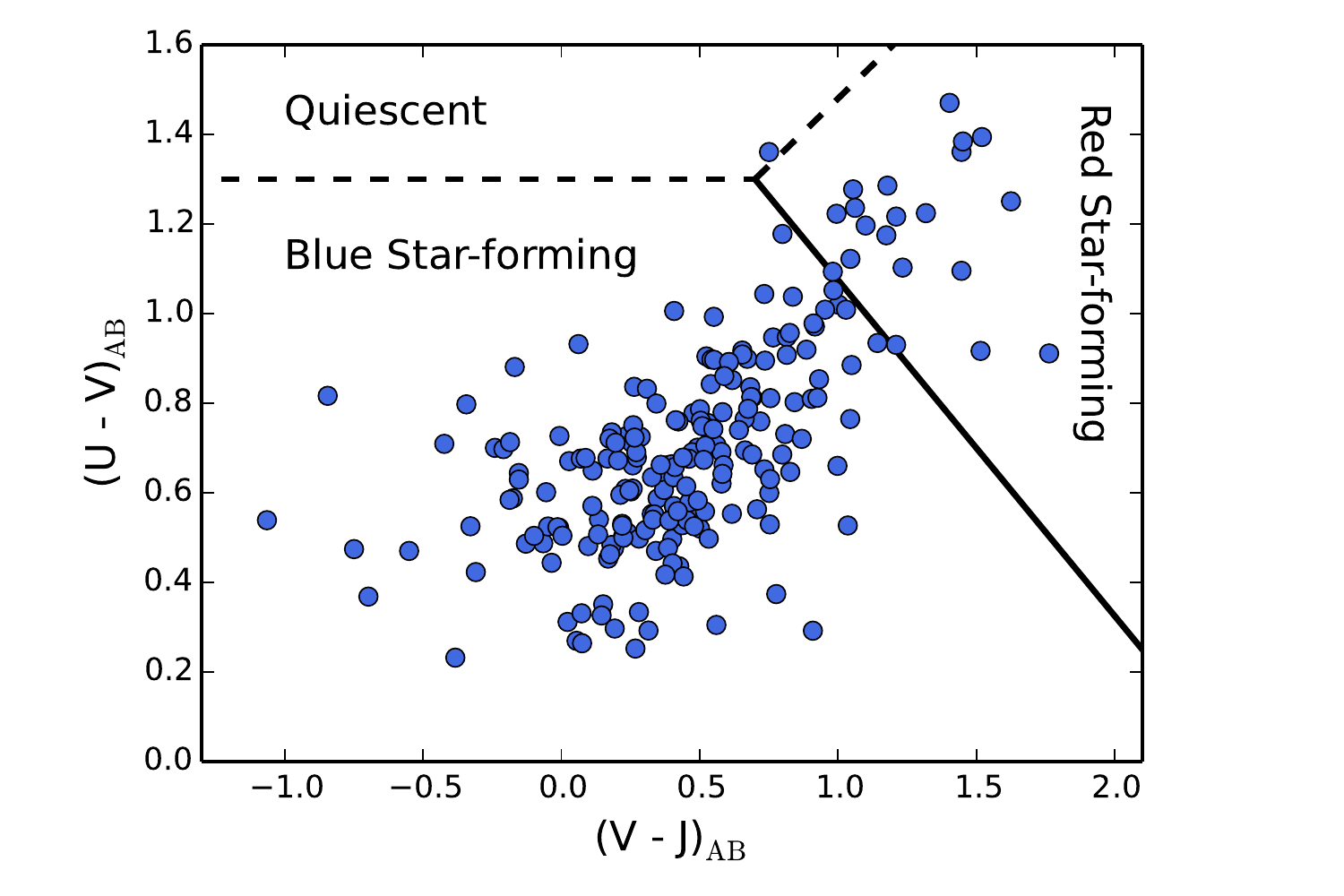} \quad
		\includegraphics[scale=.5]{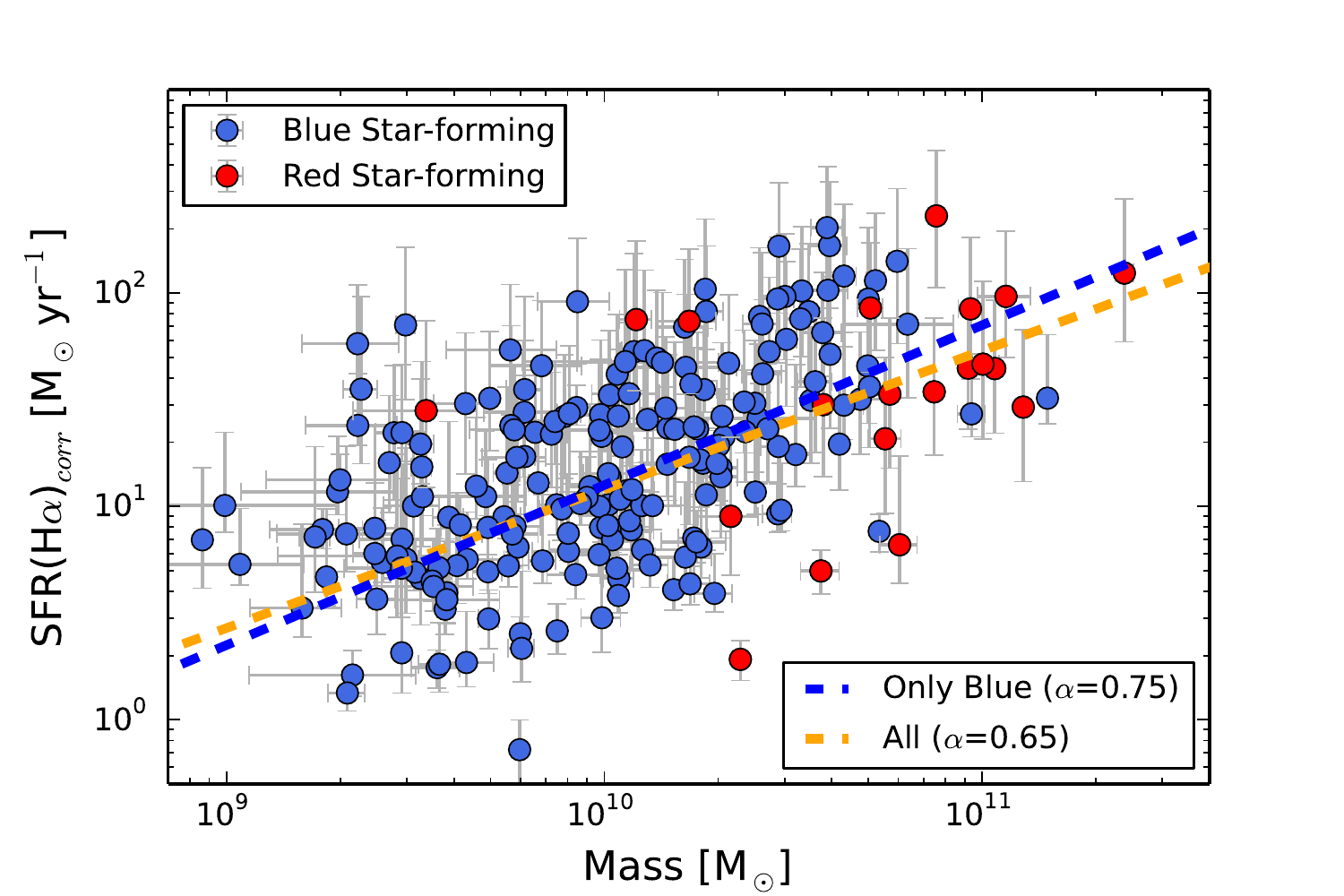}
	\end{center}
	\caption{ {\em Top:} distribution of our {\halpha} and {\hbeta} detected sample at $z = 1.4-2.6$ in the rest-frame UVJ diagram. The solid line separates blue and red star-forming galaxies.
	{\em Bottom:} SFR({\halpha}) versus $M_*$ for the blue and red star-forming galaxies. The log(SFR)-log($M_*$) slope of the blue star-forming sample is steeper than the slope of the whole sample, including the red star-forming galaxies.
	}
	\label{fig:uvj}
\end{figure}

\subsection{Sample Selection}
\label{sec:slope_sample}
Selection effects can have a major effect on the derived log(SFR)-log($M_*$) slope.
The samples that are selected based on rest-frame UV colors (such as Lyman-break selection, \citealt{steidel03,steidel04}) are biased towards less dusty galaxies.
Adding the redder and more dust-attenuated {\em star-forming} galaxies to these samples results in shallower slopes \citep{whitaker12b,speagle14}. To test this effect on our sample, we divided the star-forming galaxies into a blue bin and red bin, based on their rest-frame UV and VJ colors. Previously, we used the UVJ diagram to remove the quiescent galaxies from our sample (Section~\ref{sec:hasample}), and now we set another criterion on the star-forming part of the diagram to separate the blue and red star-forming galaxies. The selection regions are shown in Figure~\ref{fig:uvj}. After the removal 21 galaxies that were classified as red star-forming, the slope of the relation increases to $0.75\,\pm\,0.11$. 
These galaxies are both more dust-attenuated and older compared to the rest of the sample.
A steeper slope is expected as the red star-forming galaxies dominate the lower right corner of the main sequence (those with lower specific SFRs). In conclusion, the red star-forming galaxies should be included in a complete sample, otherwise the slope of the log(SFR)-log($M_*$) relation may be overestimated.

Considering that the MOSDEF sample is a near-IR selected sample and because the {\halpha} luminosity is less affected by dust attenuation than the shorter-wavelength UV luminosity, this study is less prone to the bias associated with UV selection.

\begin{figure}[tbp]
	\begin{center}
		\includegraphics[scale=.6]{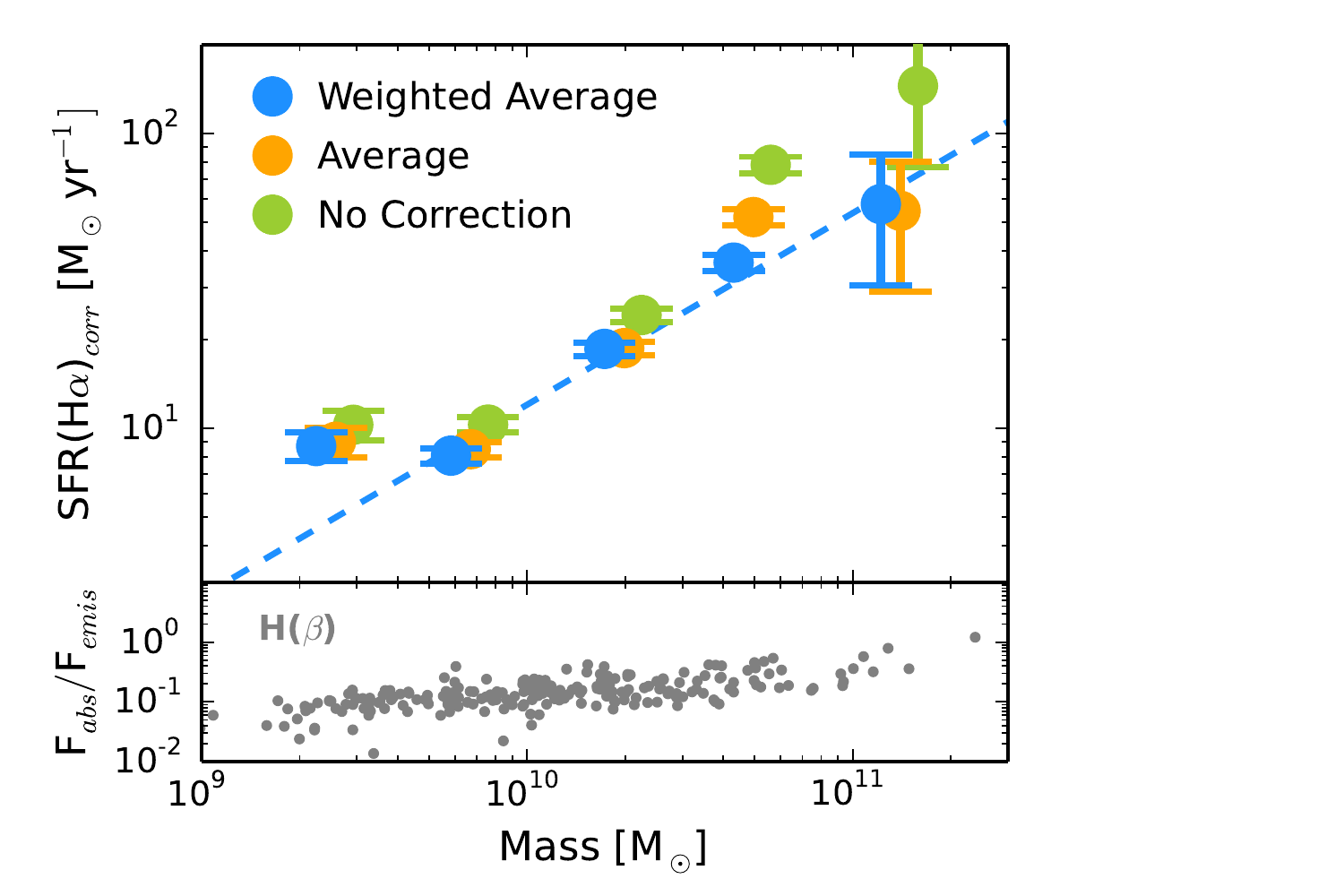}
	\end{center}
	\caption{{\em Top:} the SFR-$M_*$ relation for stacks of galaxies in five bins of stellar mass, the difference being in how the stacked {\halpha} and {\hbeta} luminosities are corrected for the underlying Balmer absorption. For illustrative purposes each set of colored points is slightly shifted in mass direction. {\em Blue symbols:} the Balmer absorption in each bin is estimated as the weighted average of the individual absorption fluxes. {\em Orange symbols:} the same as the blue points, but the average is used instead of the weighted average. {\em Green symbols:} No absorption correction is applied.
{\em Bottom:} ratio of the {\hbeta} absorption flux to the emission flux as a function of stellar mass. The absorption flux becomes stronger with stellar mass, which results in a more significant correction for galaxies with higher stellar masses.
	}
	\label{fig:balmabs}
\end{figure}

\subsection{Balmer Absorption Correction}
Another factor that affects the SFR-$M_*$ parameterization in studies that use the hydrogen Balmer lines is the Balmer absorption correction. For the star-forming galaxies considered in this study, the difference between the absorption corrected SFRs and those that are not corrected becomes larger at higher masses because the absorption flux, which is produced in the atmosphere of primarily A-type stars, increases as a function of stellar mass (see the bottom panel in Figure~\ref{fig:balmabs}). The galaxies with higher stellar masses have more mature stellar populations and hence higher Balmer absorption equivalent widths.

In Figure~\ref{fig:balmabs}, we show the stacks of SFR({\halpha}) in bins of stellar mass, corrected for the Balmer absorption in two different ways as well as the stacks without any corrections.
The absorption correction affects both the value of the observed SFR({\halpha}) and the dust correction (i.e., the ratio of {\halpha} to {\hbeta}). We used the following correction method in this study. In each stellar mass bin, the Balmer absorption of individual galaxies inferred from the stellar population models were averaged, weighted by the inverse variance calculated from the errors in the individual Balmer emission line luminosities, and added to the stacked emission line luminosities. These are blue symbols in Figure~\ref{fig:balmabs}.

On the other hand, as presented in Figure~\ref{fig:balmabs}, not correcting for the Balmer absorption (green symbols) results in values that are inconsistent with the blue linear fit by as large as $9\sigma$ at $\log(M_*) \sim 10.6$. The absorption corrections reduce the inferred dust-corrected SFRs by values that are typically larger at larger masses. (Here, by factors of 1.3, 2.1, and 2.5 at the stellar masses of $\sim 10^{10.2}$, $10^{10.6}$, $10^{11.1} \msun$.) As a result, the SFRs at higher masses are lower once the Balmer absorption correction is applied. In other words, not correcting for the underlying Balmer absorption results in higher SFRs at high masses and hence, a steeper slope.

In conclusion, we emphasize the importance of accounting for the Balmer absorption when using the Balmer emission lines to quantify the SFR-$M_*$ relation. Using the Balmer lines that are not corrected for the underlying absorption results in higher SFR values, especially at the large masses where the Balmer absorption is stronger, and affects the parameterization of the SFR-$M_*$ relation.
The stacked points in bins of stellar mass, for which the weighted averaged Balmer absorption corrections are applied, are in the best agreement (within $\sim 1-2\sigma$) with the $\log(\text{SFR})-\log(M_*)$ linear regression fitted to the individual galaxies (the blue line in Figure~\ref{fig:balmabs}).

\begin{figure}[tbp]
	\begin{center}
		\includegraphics[scale=.6]{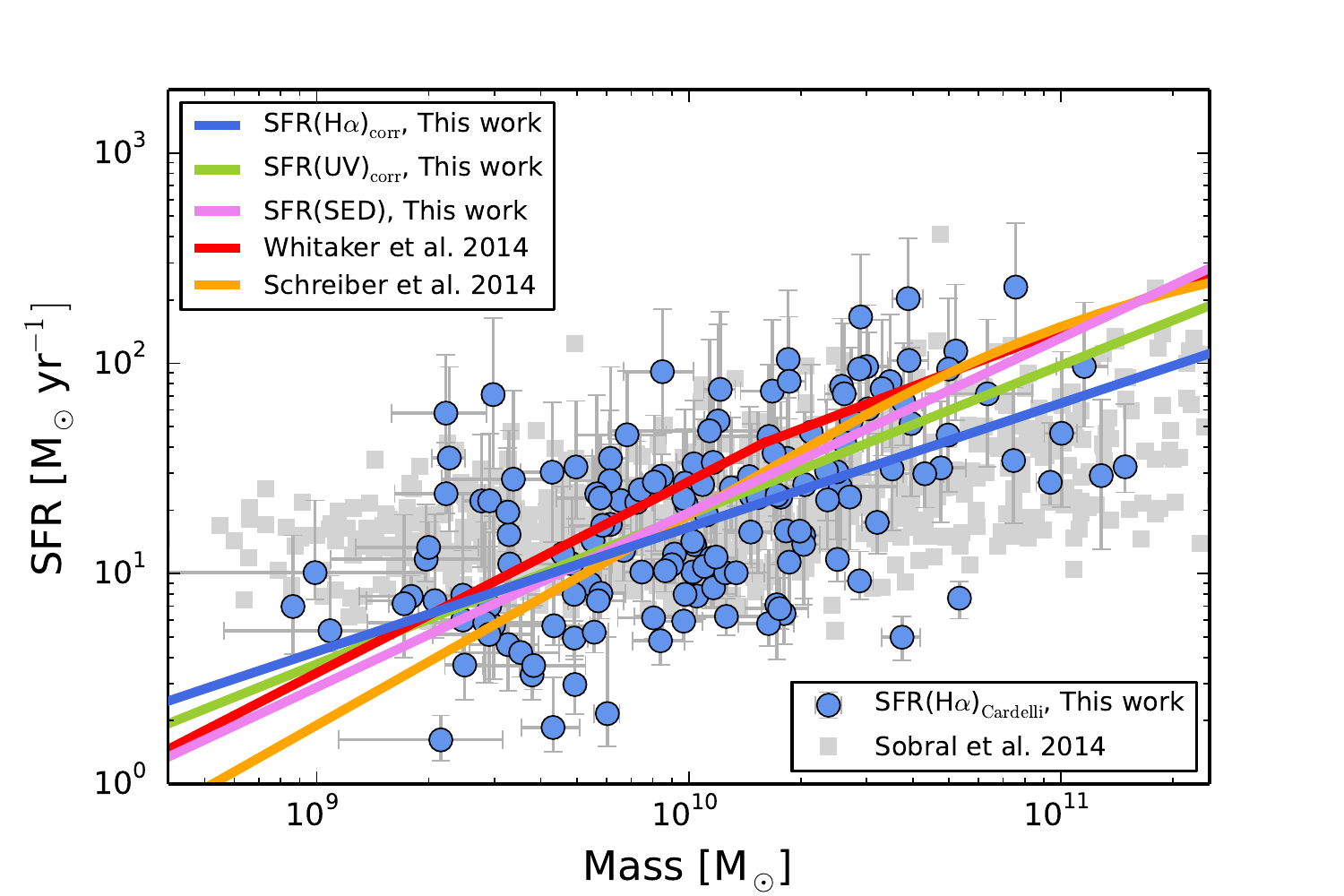}
	\end{center}
	\caption{ The comparison of SFR-$M_*$ relation for various studies. The blue points and the blue line are our $z = 2.09-2.61$ results, using SFR({\halpha}) corrected for dust according to the Balmer decrement. The gray squares are from \citet{sobral14} narrow-band selected star-forming sample, where they used SFR({\halpha}) dust corrected based on the local mass-dependent correction relation of \citet{garnbest10}. The \citet{whitaker14b} and \citet{schreiber14} relations are based on SFR(UV)+SFR(IR). We also plotted our results based on the SFR(UV) and SFR(SED) values.
	The relations from the literature are converted to a Chabrier IMF, when necessary.
	}
	\label{fig:compare}
\end{figure}

\section{Comparison to Other Studies}
\label{sec:compare}

The SFR-$M_*$ relation has been studied extensively. The parameterization of the SFR-$M_*$ relation depends on the SFR indicator and the typical timescale over which it traces star formation. First we consider studies that also use {\halpha} to quantify SFRs and therefore provide the most direct comparison. \citet{koyama13}, \citet{sobral14}, and \citet{darvish14} used {\halpha} narrow-band selected samples at $z\sim 2$. Such studies, however, tend to be biased towards galaxies with larger {\halpha} equivalent widths. These three studies also did not measure {\hbeta}, and relied on a mass-dependent dust correction from \citet{garnbest10}. The \citet{garnbest10} relation was derived based on a local sample of galaxies but its validity has not been verified for $z\sim 2$ galaxies.
The gray symbols in Figure~\ref{fig:compare} are the data points from \citet{sobral14}. At the low mass end, $M_* \lesssim 10^{10}\msun$, the high equivalent width bias mentioned above and the effect of the improper dust correction are evident. When we applied an average mass-dependent dust correction based on our Balmer decrement measurements to the \citet{sobral14} observed SFRs, we see a better agreement with our results at low masses. The \citet{sobral14} main sequence is tighter compared to ours, mainly because the scatter in the dust extinction--$M_*$ relation is not taken into account.

\citet{atek14} is another {\halpha} study, where they used {\em HST}/WFC3 near-IR grism spectroscopy for a sample of emission line galaxies at $0.3\lesssim z\lesssim 2.5$. They used the {\halpha}, {\hbeta}, [O{\sc ii}]$\lambda$3727, and [O{\sc iii}]$\lambda$5008 nebular emission lines as the SFR indicators, dust corrected in bins of stellar mass. The log(SFR)-log($M_*$) slope they found was $\sim 0.4$ at $z\sim 2.3$, though they also identified a Malmquist bias as being a potential issue in their analysis.
The SFR-$M_*$ relation was also investigated by \citet{erb06c} using a sample 114 star-forming galaxies at $z\sim 2$ with {\halpha} spectra corrected for dust extinction by the SED {\ebmv} and the Calzetti curve. \citet{zahid12} used the \citet{erb06c} sample to fit a linear relation to the median SFRs in bins of stellar masses. Similar to the other {\halpha} studies, they also found a shallow slope of $0.46\,\pm\,0.07$.

We also considered SFR-$M_*$ studies where the total SFR was estimated as the sum of UV and IR-based SFRs, plotting the SFR-$M_*$ relations of \citet{whitaker14b} and \citet{schreiber14} on top of our points in Figure~\ref{fig:compare}. The log(SFR)-log($M_*$) slopes of these studies are generally steeper than our slope. As we used the Balmer decrement to correct the observed SFR({\halpha}) for dust extinction, as opposed to using the IR data, it is probable that we missed optically thick star-forming regions, particularly in massive galaxies with high specific SFRs. This bias may be the reason of our shallower slope compared to the studies with the IR data.
There are several important differences between the \citet{whitaker14b} and \citet{schreiber14} studies and ours, such as the samples, the SFR diagnostics, and the statistical methods used for the fit. At this point, it is not trivial to reconcile these discrepancies and discuss the agreement or disagreement between these studies and ours. A more informative comparison will become possible once we incorporate IR data into our analysis.

A few studies found indications of a non-linear SFR-$M_*$ relation \citep[e.g.,][]{schreiber14,whitaker14b,lee15}. They suggested that the log(SFR)-log($M_*$) slope flattens at high masses so that a single power law cannot explain the SFR-$M_*$ relation. According to these studies, at higher redshifts the high-mass flattening becomes less prominent and at $z\sim 2$ the turning point shifts to $M_*\sim 10^{10.5}$. 
A turning point in the SFR-$M_*$ relation is not obvious from the individual detections in our sample, nor from the stacks (Figure~\ref{fig:stack}). However, because of the large uncertainty in the most massive stacked bin of our galaxies (see the blue and orange points in Figure~\ref{fig:balmabs}), we can not confidently rule out the flattening at the high-mass end.

\section{Conclusions and Discussion}
\label{sec:summary}

As part of the MOSDEF survey, we used a sample of 185 star-forming galaxies with {\halpha} and {\hbeta} spectroscopy at $M_* > 10^{9.5}\msun$ to study the star-forming main sequence relation at $z\sim 2$. The parent catalog is {\em H}-band (i.e., rest-frame optically) selected and is accompanied by ancillary multi-band photometry \citep{skelton14}, from which the UV-based SFRs were measured. The stellar masses were derived by comparing the emission-line corrected photometric SEDs with stellar population models.

After taking into account the measurement uncertainties in the SFR and stellar mass, we found a measurement-subtracted scatter of 0.31\,dex in the log(SFR({\halpha}))-log($M_*$) relation at $2.09<\,z<\,2.61$, which is 0.06\,dex larger than what we found based on the UV SFRs (see, Table~\ref{tab:param}). Although in theory the time-scale variations in SFR can be traced by using different SFR diagnostics, such as UV and {\halpha}, in the SFR-$M_*$ relation \citep{sparre15}, we argue that in the absence of direct measurements of galaxy-to-galaxy variations in the attenuation curves and the IMF, the difference in the scatter of the log(SFR({\halpha}))-log($M_*$) relation and the scatter of the log(SFR(UV))-log($M_*$) relation can not be used to constrain the stochasticity of star formation. Given these variations, the SFR(UV) scatter could be as large or greater than the SFR({\halpha}) scatter. Theorists trying to reproduce the observed scatter in the log(SFR)-log($M_*$) relation should proceed with caution, based on the results shown in this paper.

The scatter in the log(SFR)-log($M_*$) relation is tighter when the SFR(UV) is corrected for extinction by {\ebmv} derived from the SED modeling or when the SED-inferred SFRs are used, because the UV slope regulates the color-excess in the SED fitting from which the stellar masses were derived.

In addition to the variations in the dust attenuation, IMF, and metallicity, the scatter in the log(SFR)-log($M_*$) relation may be due to other variables such as geometry, size, SFR surface density, and age \citep[e.g.,][]{wuyts13,hemmati14}. The significance of the effect of these variables on the scatter will be investigated in future studies.
Although in agreement with some simulations \citep{torrey14,sparre15}, the measurement-subtracted scatter we found is larger than some of the other works \citep[e.g.,][]{dutton10a,dominguez15}. The discrepancy may indicate that the galaxy accretion and feedback processes are not as smooth as predicted by the simulations. 

We found a constant slope of $0.58\,\pm\,0.10$ for the $z = 2.1 - 2.6$ sample and $0.65\,\pm\,0.08$ when we used the whole sample at $z= 1.4 - 2.6$, for $M_* \ge 10^{9.5}\msun$. 
Galaxy evolution simulations generally find a close-to-unity slope for the log(SFR)-log($M_*$) relation \citep[e.g.,][]{dutton10a,torrey14,sparre15}, which translates to a stellar mass-independent specific SFR (i.e., SFR/$M_*$). Our shallower slope indicates higher (lower) SFR at low (high) stellar masses. As the slope of the log(SFR)-log($M_*$) represents the star-formation efficiency, one explanation for close-to-unity slope in the galaxy simulations is that star formation is too inefficient at the low stellar-mass end due to feedback processes. The steep slope in the simulations may also be caused by a lack of quenching feedback that causes a dearth of red star-forming galaxies at the high-mass end of the SFR-$M_*$ relation.

The slope of the log(SFR)-log($M_*$) is influenced by various observational and measurement effects that could explain discrepancies in the slopes reported in different observational studies. We demonstrated three of the main effects here: dust correction, sample biases, and stellar absorption correction (the last one is specific to the emission line studies). Correcting the {\halpha} emission line with the SED inferred $E(B-V)_{\rm SED} \times 2.27$, with the assumption of the Calzetti curve, leads to a log(SFR)-log($M_*$) slope of close to unity. A bias against dusty galaxies with low specific SFRs, as might be the case for UV-selected samples, may also result in steeper slopes of the log(SFR)-log($M_*$) relation. Finally, we showed that the Balmer absorption correction has a significant effect on the dust-corrected SFRs and not applying the correction to the Balmer emission lines could lead to an overestimation of the log(SFR)-log($M_*$) slope.

The evolution of the slope, normalization, and the scatter of the log(SFR)-log($M_*$) relation is essential to our understanding of the processes that govern galaxy growth during cosmic time. Once the whole MOSDEF sample becomes available, we have access to a large sample of rest-frame optical spectra for galaxies at $1.4 \lesssim z\lesssim 3.8$, which makes it possible to study the SFR-$M_*$ evolution in redshift, using instantaneous and accurately-measured SFRs in a representative sample. Furthermore, incorporating IR data into our analysis will allow us for a more detailed comparison with the existing studies.

~\\
We thank the referee for the constructive comments.
Funding for the MOSDEF survey is provided by NSF AAG grants AST-1312780, 1312547, 1312764, and 1313171 and grant AR-13907, provided by NASA through a grant from the Space Telescope Science Institute. The authors thank the MOSFIRE instrument team for building this powerful instrument, and for taking data for us during their commissioning runs. We are grateful to Marc Kassis at the Keck Observatory for his many valuable contributions to the execution of this survey. We also acknowledge the 3D-HST collaboration, who provided us with spectroscopic and photometric catalogs used to select MOSDEF targets and derive stellar population parameters.
IS thanks Shoubaneh Hemmati and Mostafa Khezri for useful discussions and feedback on the manuscript. We thank David Sobral for providing the HiZELS {\halpha} catalog that was used as a comparison to our results, and James Aird for his help in matching AGN selection catalogs.
Support for IS is provided through the National Science Foundation Graduate Research Fellowship DGE-1326120. NAR is supported by an Alfred P. Sloan Research Fellowship. MK acknowledges support from the Hellman Fellows fund. ALC acknowledges funding from NSF CAREER grant AST-1055081. We wish to extend special thanks to those of Hawaiian ancestry on whose sacred mountain we are privileged to be guests. Without their generous hospitality, most of the observations presented herein would not have been possible. This work is also based on observations made with the NASA/ESA Hubble Space Telescope (programs 12177, 12328, 12060--12064, 12440--12445, 13056), which is operated by the Association of the Universities for Research in Astronomy, Inc., under NASA contract NAS 5-26555.

\bibliographystyle{apj}

\end{document}